\documentclass[aps,twocolumn,showpacs,dvips]{revtex4}
\usepackage{feynmp}
\usepackage{amssymb}
\usepackage{epsfig}
\usepackage{graphicx}
\usepackage{subfigure}
\usepackage{hyperref}
\usepackage{bbm}

\begin{document}

\title{Two-loop disorder effects on the nematic quantum criticality in $d$-wave superconductors}

\author{Jing Wang\footnote[1]{jwang315@ustc.edu.cn}}

\affiliation{Department of Modern Physics, University of Science and
Technology of China, Hefei, Anhui, 230026, P.R. China}

\begin{abstract}
The gapless nodal fermions exhibit non-Fermi liquid behaviors at the
nematic quantum critical point that is supposed to exist in some
$d$-wave cuprate superconductors. This non-Fermi liquid state may be
turned into a disorder-dominated diffusive metal if the fermions
also couple to a disordered potential that generates a relevant
perturbation in the sense of renormalization group theory. It is therefore
necessary to examine whether a specific disorder is relevant or not. We
study the interplay between critical nematic fluctuation and random chemical
potential by performing renormalization group analysis. The parameter that
characterizes the strength of random chemical potential is marginal
at the one-loop level, but becomes marginally relevant after
including the two-loop corrections. Thus even weak random chemical
potential leads to diffusive motion of nodal fermions and the significantly
critical behaviors of physical implications, since the strength flows
eventually to large values at low energies.
\end{abstract}


\maketitle


\section{Introduction}

The low-lying elementary excitations of $d$-wave cuprate superconductors are
known to be massless nodal fermions that have a linear dispersion and fulfill
the relativistic Dirac equation . These fermions govern many of the unusual
low temperature properties of the superconducting state. In the past
twenty years, there have been extensive experimental signatures
\cite{Orenstein_Millis2000Science} supporting the fact that the
nodal fermions are nearly non-interacting and have a rather long
lifetime. However, this feature can be significantly changed if the
fermions interact with certain critical bosonic mode. Recently,
various measurements observed an anisotropy in the physical
properties of some cuprate superconductors \cite{Vojta2009AP,
Ando2002PRL, Hinkov2008Science, Daou2010Nature, Lawler2010Nature,
Borzi2007Science, Chuang2010Science}. Such an anisotropy is usually
attributed to the emergence of a novel electronic nematic order
\cite{Kivelson1998Nature,Yamase2000JPSJ,Halboth2000PRL,Fradkin2010ARCMP},
which spontaneously breaks $C_4$ symmetry of the system down to $C_2$ symmetry.
Both experimental~\cite{Hinkov2008Science,Daou2010Nature,Lawler2010Nature} and
theoretical~\cite{Vojta2009AP,Sachdevbook,Yamase2007PRB,Zhang2002PRB,Raghu2009PRB,
Moon2010PRB} studies suggest that a zero temperature nematic
quantum critical point is therefore expected to exist somewhere in the superconducting dome.
In the vicinity of this point, the nematic order
parameter fluctuates quantum-mechanically around its vanishing mean
value. The critical nematic fluctuation couples strongly to the
nodal fermions, which gives rise to severe fermion damping \cite{Vojta2000PRL,
Vojta2000PRB, Vojta2000IJMPB, Khveshchenko2001PRL, Kim2008PRB} and other
striking properties \cite{Oganesyan2001PRB, Huh2008PRB, Xu2008PRB,
Fritz2009PRB, Liu2012PRB, Wang_Liu2013NJP, She2014, Liu2015}.

The quantum critical phenomena associated with the critical nematic
fluctuation become more interesting, and meanwhile more complicated,
when there are certain amount of disorders. Disorders play an
essential role in modern condensed matter physics \cite{Lee1985RMP,
Altland2002PR}, and can result in a plenty of prominent phenomena,
such as Anderson localization and metal-insulator transition. The
past two decades have witnessed intense research activities devoted
to the study of two-dimensional massless Dirac fermions moving in a
disordered potential. Realistic systems that contain these fermions
include the aforementioned $d$-wave cuprate superconductors
\cite{Sachdevbook, Altland2002PR, Lee2006RMP}, graphene
\cite{Geim2005Nature, Aleiner2006PRL, Foster2006PRL,
CastroNeto2009RMP, Dassarma2011RMP, Kotov2012RMP}, quantum Hall
effect \cite{Ludwig1994PRB, Furneaux1995PRB, Ye1998PRL_1999PRB}, and
topological insulators \cite{Kane2010RMP}. In systems composed of
Dirac fermions, there can be three sorts of disorders
\cite{Nersesyan1995NPB}. According to the coupling with nodal
fermions, the disorders might be random chemical potential, random
mass, or random gauge potential \cite{Stauber2005PRB}. Theoretical
analysis has demonstrated that these disorders can produce
different behaviors of Dirac fermions \cite{Sachdevbook,
Altland2002PR, Lee2006RMP, Geim2005Nature, Aleiner2006PRL,
Foster2006PRL, CastroNeto2009RMP, Dassarma2011RMP, Kotov2012RMP,
Ludwig1994PRB, Furneaux1995PRB, Ye1998PRL_1999PRB, Kane2010RMP}.

Back to the $d$-wave cuprate superconductors, an important question
is whether these disorders drive an instability of the nematic
quantum critical point. It is also interesting to examine the
influence of disorders on the low-energy properties of nodal
fermions. Recently, we have studied this problem by means of
renormalization group (RG) techniques at the one-loop level
\cite{WLK2011PRB, Wang2013PRB}. In particular, we have introduced a
parameter $\zeta$ to characterize the effective strength of
disorder, and calculated the RG flows of $\zeta$ after taking into
account the interplay between fermion-nematic interaction and
fermion-disorder interaction. It was found that $\zeta$ vanishes in
the low-energy region in the presence of random mass and random
gauge potential \cite{WLK2011PRB}. However, $\zeta$ does not flow at
all in the case of random chemical potential, i.e., $d\zeta/dl = 0$
with $l$ being a running length scale \cite{WLK2011PRB}. Therefore,
the parameter $\zeta$ for random chemical potential is considered as
marginal. In the other two cases, $d\zeta/dl < 0$, so $\zeta$
vanishes in the low-energy regime and is hence irrelevant. If the
perturbative expansion is reliable, the result that $\zeta$ vanishes
in the low-energy regime obtained in the cases of random mass and
random gauge potential will not be changed by higher order
corrections. Nevertheless, the same conclusion cannot be simply
reached in the case of random chemical potential.

In the spirit of RG theory  \cite{Wilson1975RMP, Polchinski,
Shankar1994RMP}, the marginal nature of certain parameter is hardly
stable against higher order corrections. It often happens that a
marginal parameter is turned to marginally relevant or marginally
irrelevant once higher order corrections are taken into account.
These two fates of disorder parameter may induce different behaviors
of physical quantities. In order to address this issue, we need to
go beyond the one-loop RG analysis, and make detailed two-loop (or
even higher order) calculations. Interestingly, a recent work
\cite{Roy2014PRB} has revealed that two-loop corrections are able to
make an important contribution to the quantum criticality of some
particular three-dimensional Dirac semimetals, which partly
motivated the present work.

In this paper, we combine the $1/N$-expansion method and the replica
technique to analyze the interplay of nematic fluctuation and
disorder scattering (in the following text we use "disorder" to
uniquely denote random chemical potential). The perturbation
expansion is carried out up to two-loop level. We derive a number of
RG flow equations for fermion velocities $v_F$ and $v_\Delta$,
velocity ratio $v_\Delta/v_F$, and disorder parameter $\zeta$. These
equations are coupled to each other, and thus need to be solved
self-consistently. We show that the parameter $\zeta$ for random
chemical potential becomes marginally relevant due to two-loop
corrections. Therefore, the interaction between nodal fermions and
random chemical potential is always in the strong coupling regime at
nematic quantum critical point, since any small parameter $\zeta$
flows eventually to infinity at the lowest energy.

The rest of the paper is organized as follows. The effective field
theory and the corresponding Feynman rules are given in
Sec.~\ref{Sec_eff_theory}. We present the RG transformations and
perform detailed computations of one-loop correction to fermion
self-energy and vertex functions in Sec.~\ref{Sec_one_loop}, which
is followed by two-loop calculations in Sec.~\ref{Sec_two_loop}. In
Sec.~\ref{Sec_numerical_discussions}, we solve the RG equations and
discuss the physical implications in Sec.~\ref{Sec_diagram implications}.
Finally, we briefly summarize our results in Sec.~\ref{Sec_summary}.

\begin{figure}
\centering
\includegraphics[width=3.3in]{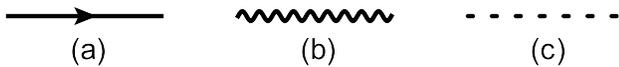}
\vspace{-0.10cm} \caption{Propagators: (a) fermion, (b) nematic, and
(c) disorder.}\label{Fig_field_propagators}
\end{figure}

\section{Effective field theory}\label{Sec_eff_theory}

The interaction between massless nodal fermions and nematic order
parameter is described by the following action \cite{Kim2008PRB,
Huh2008PRB}
\begin{eqnarray}
S &=& S_{\psi} + S_{\phi} + S_{\psi\phi},
\end{eqnarray}
where $S_{\psi}$ is the free action for nodal fermions
\begin{eqnarray}
\hspace{-0.35cm}S_{\psi} \!\!&=&\!\! \int\!\!\frac{d^{2}\mathbf{k}}{(2\pi)^{2}}
\frac{d\omega}{2\pi} \psi^{\dagger}_{1a}(-i\omega +
v_{F}k_{x}\tau^{z} + v_{\Delta}k_{y}\tau^x)\psi_{1a} \nonumber \\
\hspace{-0.35cm}&& \!\!+ \!\!\int\!\!\frac{d^{2}\mathbf{k}}{(2\pi)^2}
\frac{d\omega}{2\pi}\psi^{\dagger}_{2a} (-i\omega +
v_{F}k_{y}\tau^{z} + v_{\Delta}k_{x}\tau^{x})\psi_{2a},
\end{eqnarray}
with $\tau^{(x,y,z)}$ being the standard Pauli matrices. The linear
dispersion of nodal fermions originates from the $d_{x^2 -
y^2}$-wave symmetry of the superconducting gap of cuprates. Here,
spinor $\psi^{\dagger}_{1}$ represents the fermionic quasiparticles (QPs)
excited from nodal points $(\frac{\pi}{2},\frac{\pi}{2})$ and
$(-\frac{\pi}{2},-\frac{\pi}{2})$, and $\psi^{\dagger}_{2}$ the
other two nodal points \cite{Vojta2000PRL, Vojta2000PRB, Vojta2000IJMPB}.
The repeated spin index $a$ is summed from $1$ to $N_f$, which is the number
of fermion spin components. The ratio between Fermi velocity $v_{F}$ and gap
velocity $v_{\Delta}$ is roughly $v_{\Delta}/v_{F} \approx 1/10$,
which is determined by experiments
\cite{Orenstein_Millis2000Science}.

The action $S_{\phi}$ describes the Ising type nematic order
parameter, which is expanded in real space as
\begin{eqnarray}
S_{\phi} = \!\int \!d^2\mathbf{x}d\tau
\!\left[\!\frac{1}{2}(\partial\tau\phi)^2 + \frac{c^2}{2}(\nabla\phi)^2
+ \frac{r}{2}\phi^2 + \frac{u_0}{24}\phi^4\!\right],
\end{eqnarray}
where $\tau$ is imaginary time and $c$ is velocity for $\phi$. The
mass parameter $r$ tunes the nematic phase transition, and $r=0$
defines the corresponding quantum critical point. Parameter $u_0$ is
the strength of quartic self-interaction. The nematic order
parameter couples to nodal fermions through the simple Yukawa term
\begin{eqnarray}
S_{\psi\phi} = \int d^2\mathbf{x}d\tau\{\lambda_0
\phi(\psi^{\dagger}_{1a} \tau^{x}\psi_{1a}+
\psi^{\dagger}_{2a}\tau^{x}\psi_{2a})\}.
\end{eqnarray}

\begin{figure}
\centering
\includegraphics[width=1.99in]{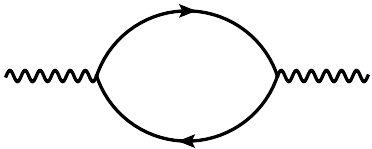}
\vspace{-0.10cm}
\caption{The polarization function for nematic order parameter.}
\label{Fig_fermion_polarization}
\end{figure}

The strong interaction between nodal fermions and nematic order can
be handled by performing $1/N_f$-expansion. The inverse of the free
propagator of $\phi$ behaves as $q^2+r$. After taking into account
the polarization function, an additional linear $q$-term will be
generated, namely $\Pi(q)\propto q$. At the low-energy regime, the
$q$-term dominates over the $q^2$-term, which then can be neglected.
Near the quantum critical point, we keep only the mass term and make
the replacement that $\phi \longrightarrow \phi/\lambda_0$ and $r
\longrightarrow N_f r \lambda^2_0$, leading to
\begin{eqnarray}
S = S_{\psi}\!+\!\!\!\int \!\!d^2\mathbf{x}d\tau\!\left\{\!\frac{N_fr}{2}
\phi^2\!+\!\phi[\psi^{\dagger}_{1a}\tau^x\psi_{1a}\!+\!
\psi^{\dagger}_{2a}\tau^x\psi_{2a}]\!\right\}\!.
\end{eqnarray}
After integrating out fermion degrees of freedom, the effective
action for $\phi$ becomes
\begin{eqnarray}
\frac{S_{\phi}}{N} = \!\frac{1}{2}\int\!\!\frac{d^2\mathbf{q}}{(2\pi)^2}\frac{d\epsilon}{2\pi}
\left[r+\Pi(\mathbf{q},\epsilon)\right]|\phi(\mathbf{q},\epsilon)|^{2}+\mathcal{O}(\phi^{4}).
\end{eqnarray}
The lowest-order Feynman diagram for the polarization function is
shown in Fig. \ref{Fig_fermion_polarization} and symbolizes the
integral
\begin{eqnarray}
\Pi(\mathbf{q},\epsilon)
=\!\!\int\frac{d^{2}\mathbf{k}}{(2\pi)^{2}}\frac{d\omega}{2\pi}
\mathrm{Tr}[\tau^{x}G^{0}_{\psi}(\mathbf{k},\omega)\tau^{x}
G^{0}_{\psi}(\mathbf{k+q},\omega+\epsilon)],\nonumber
\end{eqnarray}
where the free fermion propagator is
\begin{eqnarray}
G^{0}_{\psi}(\mathbf{k},\omega)=\frac{1}{-i\omega
+v_{F}k_{x}\tau^{z}+v_{\Delta}k_{y}\tau^{x}}.
\end{eqnarray}
As shown previously~\cite{Huh2008PRB}, the propagator for the
nematic order parameter is given by
\begin{eqnarray}
G_{\phi}^{-1}(\mathbf{q},\epsilon)
&=&\Pi(\mathbf{q},\epsilon)\nonumber\\
&=&\frac{1}{16v_{F}v_{\Delta}}
\frac{(\epsilon^{2}+v_{F}^{2}q_{x}^{2})}{(\epsilon^2+
v_{F}^{2}q_{x}^{2}+v_{\Delta}^{2}q_{y}^{2})^{1/2}}\nonumber \\
&& +\frac{1}{16v_{F}v_{\Delta}}\frac{(\epsilon^{2}+v_{F}^{2}
q_{y}^{2})}{(\epsilon^{2}+v_{F}^{2}q_{y}^{2}+
v_{\Delta}^{2}q_{x}^{2})^{1/2}}
\end{eqnarray}
in the vicinity of nematic quantum critical point $r=0$. The Yukawa
interaction has been studied extensively in recent years \cite{Kim2008PRB}.
Huh and Sachdev found a stable fixed point at which the velocity ratio $v_{\Delta}/v_F$
vanishes at the lowest energy. Such an extreme anisotropy then gives rise to a
series of striking consequences, including anomalously scaling behaviors of
specific heat, local density of states, and NMR relaxation rate $1/T_1T$ \cite{Xu2008PRB},
reduction of thermal conductivity \cite{Fritz2009PRB}, additional Cooper pairing between
nodal fermions \cite{Wang_Liu2013NJP}, strong suppression of superfluid density \cite{Liu2012PRB}
and superconducting critical temperature and anomalous scaling of the penetration depth \cite{She2014}.

Now we introduce disorders into the above model. Disorders exist in
almost all realistic condensed matter systems and are known to be
responsible for many of the low temperature properties. In the
present problem, the nodal fermions may interact with three sorts of
disorders, which can be represented by the following general term
\begin{eqnarray}
S_{\mathrm{dis}}=\int d^2\mathbf{x}\psi^{\dagger}(\mathbf{x})
\Gamma\psi(\mathbf{x})A(\mathbf{x}),
\end{eqnarray}
where the matrix $\Gamma$ is $\Gamma=\mathrm{I}$ for random chemical
potential. It will be replaced by $\tau^{y}$ in the case of random mass
and $\tau^{x,z}$ random gauge field. It is traditional to assume that
$A(\mathbf{x})$ is a quenched, Gaussian white noise potential defined
by the following correlation functions
\begin{eqnarray}
\langle A(\mathbf{x})\rangle=0; \hspace{0.5cm} \langle
A(\mathbf{x}_1)A(\mathbf{x}_2)\rangle
=\zeta\delta^2(\mathbf{x}_1-\mathbf{x}_2).
\end{eqnarray}

The disordered potential $A(\mathbf{x})$ is randomly distributed in
space, and needs to be properly averaged. A commonly utilized method
to average disorders is to introduce the so-called replica trick
\cite{Edwards1975, Lee1985RMP, Lerner2003}, which states that one
can replicate the partition function $Z$ by $R$ times with $R$ being
a positive integer, $\mathcal{Z} \rightarrow \mathcal{Z}^R$. After
doing so, one obtains a useful identity
\begin{eqnarray}
\ln\mathcal{Z} = \lim_{R \rightarrow 0} \frac{\mathcal{Z}^R-1}{R}.
\end{eqnarray}
Though $R$ is initially supposed to be an integer, we can regard it
as a continuous variable and simply take the limit $R \rightarrow
0$. It is now technically feasible to average over the replicated
partition as follows
\begin{eqnarray}
\overline{\mathcal{Z}^R} = \int\mathcal{D}VP[A]\mathcal{Z}^R,
\end{eqnarray}
where
\begin{eqnarray}
P[A] = \exp\left(-\frac{1}{\zeta}\int
d^2\mathbf{x}A^2(\mathbf{x})\right).
\end{eqnarray}
We then insert the partition function $Z$ and integrating out the
random potential $A(\mathbf{x})$, and finally obtain an effective
action
\begin{eqnarray}
\overline{S_{\mathrm{eff}}} &=& S^m_{\psi} + S^m_{\phi} +
S^m_{\psi\phi} -\frac{\zeta}{4}\int d\mathbf{x}d\tau d\tau'\nonumber \\
&&\times\psi^{\dagger}_m(\mathbf{x},\tau) \psi_m(\mathbf{x},\tau)
\psi^{\dagger}_n(\mathbf{x},\tau')\psi_n(\mathbf{x},\tau'),
\end{eqnarray}
where $m,n$ are the replica indices. To proceed, it is more
convenient to rewrite the disorder-related term in the momentum
space. Therefore, we have
\begin{eqnarray}
\overline{S_{\mathrm{eff}}}
&=& S^m_{\psi} + S^m_{\phi} + S^m_{\psi\phi} \nonumber\\
&&- \frac{\zeta}{4} \int\frac{d^2\mathbf{k}d^2
\mathbf{k'}d^2\mathbf{k''}d\omega d\omega''}{(2\pi)^{6}}
\psi^{\dagger}_m(\mathbf{k},\omega)\psi_m(\mathbf{k'},\omega)\nonumber\\
&& \times \psi^{\dagger}_n (\mathbf{k''},\omega'')
\psi_n(\mathbf{k}+\mathbf{k''}-\mathbf{k'},\omega'').
\end{eqnarray}
The propagators of the replicated action are delineated in the Fig.
\ref{Fig_field_propagators}. There are two noticeable features of
this effective action: (i) the scattering of fermions by static
disorder does not exchange energy and hence the energy of nodal
fermions is conserved; (ii) the quartic interaction connects
different replica indices.

\begin{figure}
\centering
\includegraphics[width=3.30in]{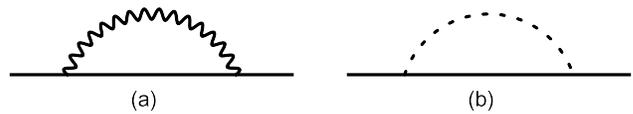}
\vspace{-0.10cm} \caption{One-loop corrections to the fermion
self-energy in the replica limit $R\rightarrow0$.}
\label{Fig_fermion_one_loop}
\end{figure}

\begin{figure}
\centering
\includegraphics[width=2.80in]{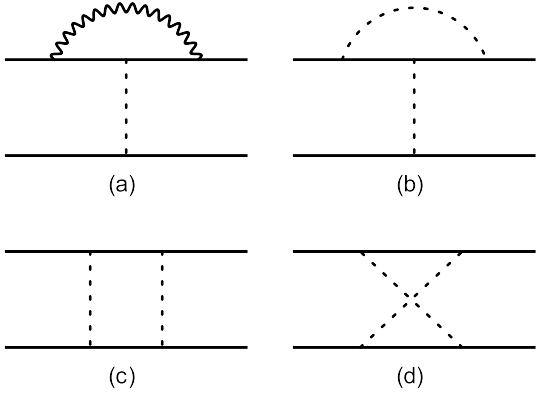}
\vspace{-0.15cm} \caption{One-loop corrections to the disorder
vertex in the replica limit $R\rightarrow0$.}\label{Fig_disorder_one_loop}
\end{figure}

We here list the one-loop and two-loop Feynman diagrams contributing
to the fermion velocities and disorder strength: (i) There are two
one-loop diagrams in the replica limit $R\rightarrow0$ which
contribute to the fermion self-energy as represented in Fig.
\ref{Fig_fermion_one_loop}; (ii) There are four one-loop diagrams in
the replica limit $R\rightarrow0$ which contribute to the disorder
vertex as represented in Fig. \ref{Fig_disorder_one_loop}; (iii)
There are seven two-loop diagrams in the replica limit
$R\rightarrow0$ which contribute to the fermion self-energy as
represented in Fig. \ref{Fig_fermion_two_loop}; (iv) There are
thirty two-loop diagrams in the replica limit $R\rightarrow0$ which
contribute to the disorder vertex as represented in Fig.
\ref{Fig_disorder_two_loop}. In the next section, we will calculate
all the associated Feynman diagrams and derive the RG flow equations
for all the physical parameters appearing in the effective action.

\section{One-loop RG analysis}\label{Sec_one_loop}

In this section, we first make RG analysis at the one-loop level.
Different from Ref.~\cite{WLK2011PRB}, the disorders are averaged
here by using the replica method. The general scheme presented in
this section will be applied directly to perform two-loop
calculations in the next section.

In order to do RG calculations, we make the following scaling
transformations \cite{Huh2008PRB, Polchinski, Shankar1994RMP},
\begin{eqnarray}
&&k_i = k'_ib,\\
&&\omega = \omega'b,\\
&&\psi_{1,2}(\mathbf{k},\omega) = \psi'_{1,2}(\mathbf{k'},\omega')
e^{\frac{1}{2}\int^{l}_{0}(4-\eta_{f})dl},\\
&&\phi(\mathbf{q},\epsilon) =
\phi'(\mathbf{q'},\epsilon')e^{\frac{1}{2}
\int^{l}_{0}(5-\eta_{b})dl},
\end{eqnarray}
where $l$ is a freely running length scale with $b=e^{-l}$ and $i=x,y$.
The scaling parameters $\eta_f$ and $\eta_b$ will be determined by the
self-energy and fermion-nematic vertex corrections. Notice that the
energy is required to scale in the same way as momentum, which means
the fermion velocities are forced to flow under RG transformations.

With the help of Dyson equation, we know that interactions induce a
self-energy correction to the free fermion propagator, i.e.,
\begin{eqnarray}
G^{-1}_{\psi}(\mathbf{k},\omega) \!&=&\! -i\omega \!+\! v_F k_x\tau^z \!+\!
v_{\Delta}k_{y}\tau^{x} \!-\! \Sigma(\mathbf{k},\omega),
\end{eqnarray}
where $\Sigma(\mathbf{k},\omega)$ is the fermion self-energy
function. In the current problem, $\Sigma(\mathbf{k},\omega)$
receives corrections from both fermion-nematic interaction and
fermion-disorder interaction. At the one-loop level, formally we
have $\Sigma(\mathbf{k},\omega) =
\Sigma^{\mathrm{L1(a)}}(\mathbf{k},\omega) +
\Sigma^{\mathrm{L1(b)}}(\mathbf{k},\omega)$, where
$\Sigma^{\mathrm{L1(a)}}$ and $\Sigma^{\mathrm{L1(b)}}$ are
generated by nematic order and disorder, respectively.

\subsection{Fermion self-energy}

The one-loop diagrams for fermion self-energy are depicted in
Fig.~\ref{Fig_fermion_one_loop}. The one-loop nematic self-energy,
shown in Fig.~\ref{Fig_fermion_one_loop}(a), has already been
calculated earlier by Huh and Sachdev \cite{Huh2008PRB}, who found
that
\begin{eqnarray}
\frac{d\Sigma^{\mathrm{L1(a)}}(\mathbf{k},\omega)}{d\ln\Lambda} \!=\!
C_1(-i\omega)+C_2v_Fk_x\tau^z+C_3v_{\Delta}k_y\tau^x,
\label{Eq_one-loop_fermion_self_energy}
\end{eqnarray}
where the coefficients $C_1$, $C_2$, and $C_3$ can be found in the
\ref{Appendix_coefficiets}. The one-loop disorder-induced fermion
self-energy $\Sigma^{\mathrm{L1(b)}}(i\omega)$, shown in
Fig.~\ref{Fig_fermion_one_loop}(b), is given by
\cite{WLK2011PRB, Wang2013PRB, Wang2013NJP}
\begin{eqnarray}
\Sigma^{\mathrm{L1(b)}}(i\omega)
&=&-2\frac{\zeta}{4}\int\frac{d^{2}\mathbf{k}}{(2\pi)^2}\Gamma
G^{0}_{\psi}(\mathbf{k},\omega)\Gamma\nonumber\\
&=&-\frac{\zeta}{4}\frac{\mathrm{factor}}{2\pi v_{F}v_{\Delta}}i\omega\ln\Lambda.
\end{eqnarray}
This expression tells us that $\Sigma^{\mathrm{L1(b)}}(i\omega)$ is
independent of the concrete form of the matrix $\Gamma$. Another
important feature is that $\Sigma^{\mathrm{L1(b)}}(i\omega)$ is
independent of momentum, which reflects the fact that the quenched
disorder is static. It is now easy to get
\begin{eqnarray}
\frac{d\Sigma^{\mathrm{L1(b)}}(i\omega)}{d\ln\Lambda} = 2C_g
i\omega,
\end{eqnarray}
where
\begin{eqnarray}
C_g=-\frac{\zeta}{4}\frac{1}{2\pi v_{F}v_{\Delta}}.
\end{eqnarray}

\subsection{Disorder vertex}

In the replica limit $R\rightarrow 0$, the one-loop diagrams for
the corrections to fermion-disorder vertex are given in
Fig.~\ref{Fig_disorder_one_loop}. The diagram of
Fig.~\ref{Fig_disorder_one_loop}(a) can be calculated by employing
the method proposed by Huh and Sachdev \cite{Huh2008PRB}. At zero
external momenta and frequency, the corresponding vertex correction
is expressed as
\begin{eqnarray}
V_{\mathrm{dis}} = \left(-\frac{\zeta}{4}\right)
\int\frac{d^{3}Q}{(2\pi)^3} H(Q)\mathcal{K}^{3}
\left(\frac{\mathbf{q}^{2}}{\Lambda^{2}}\right),
\end{eqnarray}
where $Q\equiv(\mathbf{q},\epsilon)$ is a three-momenta.  Here, $\mathcal{K}(y)$ is an arbitrary
function with $\mathcal{K}(0)=1$, and it falls off rapidly with $y$, e.g., $\mathcal{K}(y)= e^{-y}$~\cite{Huh2008PRB}. However, the results are independent of the particular choices of
$\mathcal{K}(y)$. The above equation can be converted to
\begin{eqnarray}
\frac{dV_{\mathrm{dis}}}{d\ln\Lambda} =
\left(-\frac{\zeta}{4}\right)\frac{v_{F}}{8\pi^3}
\int^{\infty}_{-\infty} dx\int^{2\pi}_{0}d\theta H(\hat{Q}),
\end{eqnarray}
where
\begin{eqnarray}
H(\hat{Q}) &=& \frac{1}{N_f}\tau^x \frac{1}{(-iv_{F}x +
v_{F}\cos\theta\tau^{z} + v_{\Delta}\sin\theta\tau^x)}
\mathrm{I}\nonumber\\
&&\times\frac{1}{(-iv_{F}x + v_{F}\cos\theta\tau^{z} +
v_{\Delta}\sin\theta\tau^{x})}\nonumber\\
&&\times\tau^{x}\frac{1}{\Pi(\hat{Q})}.
\end{eqnarray}
Here, the matrix $\mathrm{I}$ corresponds to the coupling between
nodal fermions and random chemical potential. After straightforward
computation, we have
\begin{eqnarray}
\frac{dV^{\mathrm{L1(a)}}_{\mathrm{dis}}}{d\ln\Lambda}
=2C_{5}\left(-\frac{\zeta}{4}\mathrm{I}\right),
\end{eqnarray}
where
\begin{eqnarray}
C_{5}&=&-\frac{2(v_\Delta/v_F)}{N_f\pi^3}\int^\infty_{-\infty}dx
\int^{2\pi}_{0}d\theta\nonumber \\
&&\times\frac{(x^2-\cos^{2}\theta-(v_{\Delta}/v_{F})^{2}\sin^{2}\theta)}
{(x^{2}+\cos^{2}\theta+(v_{\Delta}/v_{F})^{2}\sin^{2}\theta)^{2}}
\mathcal{G}(x,\theta)\nonumber\\
&=&-C_1.
\end{eqnarray}
Fig.~\ref{Fig_disorder_one_loop}(b) is the vertex correction due to
disorder averaging, and given by
\begin{eqnarray}
V^{\mathrm{L1(b)}}_{\mathrm{dis}}
&=&4\left(-\frac{\zeta}{4}\right)\left(-\frac{\zeta}{4}\right)\nonumber\\
&&\times\int\frac{d^{2}\mathbf{p}}{(2\pi)^{2}}\mathrm{I}G^0_{\psi}(\omega,\mathbf{p})
\mathrm{I} G^0_{\psi}(\omega,\mathbf{p+k})\mathrm{I}.
\end{eqnarray}
Taking external momentum $\mathbf{k} = 0$ and keeping only the
leading divergent term, we find that
\begin{eqnarray}
\frac{dV^{\mathrm{L1(b)}}_{\mathrm{dis}}}{d\ln\Lambda}
=4 C_{\Gamma}\left(-\frac{\zeta}{4}\mathrm{I}\right),
\end{eqnarray}
where
\begin{eqnarray}
C_{\Gamma} = -\frac{\zeta}{4}\frac{1}{2\pi v_{F}v_{\Delta}} =
C_{g}.
\end{eqnarray}
The contributions from Fig.~\ref{Fig_disorder_one_loop}(c) and
Fig.~\ref{Fig_disorder_one_loop}(d) cancel each other
\cite{Roy2014PRB}.

\subsection{RG Equations with disorder at one-loop level}

RG theory requires that the momentum shell between $b\Lambda$ and
$\Lambda$ should be integrated out, while keeping the $-i\omega$
term invariant. Using the one-loop contribution to fermion
self-energy, we have
\begin{eqnarray}
&& \int^{b\Lambda}d^{2}\mathbf{k} d\omega \psi^{\dagger}
\left[-i\omega - C_1(-i\omega)\ln\frac{\Lambda}{b\Lambda}\right.\nonumber \\
&&\left. + 2C_g(-i\omega)\ln\frac{\Lambda}{b\Lambda}\right]\psi \nonumber \\
&=& \int^{b\Lambda}d^{2}\mathbf{k} d\omega \psi^{\dagger}
(-i\omega)\left[1+(2C_g - C_1)l\right]\psi \nonumber \\
&\approx& \int^{b\Lambda}d^{2}\mathbf{k} d\omega
\psi^{\dagger}(-i\omega)e^{(2C_g - C_1)l}\psi.
\end{eqnarray}
After scaling transformation, this term goes back to the
corresponding free form, which implies that
\begin{eqnarray}
\eta_{f} = 2C_{g}-C_{1},
\end{eqnarray}
at the one-loop level. The kinetic terms should also be invariant
under scaling transformation, so we obtain the following RG
equations up to one-loop level for fermion velocities:
\begin{eqnarray}
\frac{dv_{F}}{dl} &=& (C_{1}-C_{2}-2C_{g})v_{F}, \\
\frac{dv_{\Delta}}{dl} &=& (C_{1} - C_{3} - 2C_{g})v_{\Delta}.
\end{eqnarray}
Based on these expressions, the one-loop RG equation for the ratio
between gap velocity and Fermi velocity can be derived as
\begin{eqnarray}
\frac{d\frac{v_{\Delta}}{v_{F}}}{dl}
=(C_{2}-C_{3})\frac{v_{\Delta}}{v_{F}}.
\end{eqnarray}

The disorder parameter $\zeta$ indirectly appears in the above equations which exists
in the $C_g$. Due to the interplay of nematic fluctuation and disorder scattering,
this parameter also flows under RG transformation. Including one-loop correction
due to nematic and disorder interactions yields
\begin{widetext}
\begin{eqnarray}
&&\int^{b\Lambda}d^2\mathbf{k}d^2\mathbf{k'}d^2\mathbf{k''}d\omega d\omega''
\psi^{\dagger}_m(\mathbf{k},\omega)\psi_m(\mathbf{k'},\omega)\psi^{\dagger}_n
(\mathbf{k''},\omega'')\psi_n(\mathbf{k}+\mathbf{k''}-\mathbf{k'},\omega'')
\left(-\frac{\zeta}{4}\right)\left(\mathrm{I} -
2C_{1}\mathrm{I}\ln\frac{\Lambda}{b\Lambda} +
4C_{g}\mathrm{I}\ln\frac{\Lambda}{b\Lambda}\right)\nonumber \\
&=& \int^{b\Lambda}d^2\mathbf{k}d^2\mathbf{k'}d^2\mathbf{k''}
d\omega d\omega''\psi^{\dagger}_m(\mathbf{k},\omega)
\psi_m(\mathbf{k'},\omega)\psi^{\dagger}_n(\mathbf{k''},\omega'')
\psi_n(\mathbf{k}+\mathbf{k''}-\mathbf{k'},\omega'')
\left(-\frac{\zeta}{4}\right) \mathrm{I}\left[1 + 2(2C_{g} -
C_{1})l\right] \nonumber\\
&\approx&\int^{b\Lambda}d^2\mathbf{k}d^2\mathbf{k'}d^2\mathbf{k''}
d\omega d\omega'' \psi^{\dagger}_m(\mathbf{k},\omega)\psi_m(\mathbf{k'},\omega)
\psi^{\dagger}_n(\mathbf{k''},\omega'')\psi_n(\mathbf{k}+\mathbf{k''}-\mathbf{k'},\omega'')
\left(-\frac{\zeta}{4}\right)\mathrm{I}e^{2(2C_{g}-C_{1})l}.
\end{eqnarray}
After redefining energy, momentum, and field operators, we have
\begin{eqnarray}
&&\int^{\Lambda}d^2\mathbf{k}d^2\mathbf{k'}d^2\mathbf{k''}d\omega
d\omega''\psi^{\dagger}_m(\mathbf{k},\omega)\psi_m(\mathbf{k'},\omega)
\psi^{\dagger}_n(\mathbf{k''},\omega'')\psi_n(\mathbf{k}+\mathbf{k''}-\mathbf{k'},\omega'')
\left(-\frac{\zeta}{4}\right)\mathrm{I}e^{2(2C_{g}-C_{1})l}.
\end{eqnarray}
\end{widetext}
Since $\eta_{f}=2C_{g}-C_{1}$, it is easy to obtain
the following one-loop RG equation for $\zeta$,
\begin{eqnarray}
\frac{d\zeta}{dl}=0.
\end{eqnarray}
From this result, we know that the strength parameter for random
chemical potential $\zeta$ is marginal at the one-loop level.

\section{Two-loop RG analysis}\label{Sec_two_loop}

As shown previously in Ref.~\cite{WLK2011PRB} and also in the last
section, we have found that the strength parameter of random
chemical potential is marginal at the one-loop level near the
nematic quantum critical point located in a $d$-superconductor.
According to the RG theory, the fate of one marginal parameter can
be changed once higher order corrections are incorporated.
Specifically, the marginal parameter may be turned to either
relevant or irrelevant by the corrections. Recent work found that
the two-loop corrections do make a significant contribution to the
quantum criticality of three-dimensional Dirac semimetals
\cite{Roy2014PRB}. It is therefore important to examine the fate of
marginal random chemical potential against higher order corrections.
To address this issue, we now go beyond the one-loop analysis given
in Ref.~\cite{WLK2011PRB} and perform a detailed two-loop RG
analysis.

The Feynman diagrams at two-loop level are presented in
Fig.~\ref{Fig_fermion_two_loop} and
Fig.~\ref{Fig_disorder_two_loop}. These diagrams can by calculated
by employing the methods introduced in \cite{Huh2008PRB},
\cite{Mishchenko2007PRL}, \cite{Vafek2008PRB} and \cite{Roy2014PRB}.
Paralleling the procedures used in Sec.~\ref{Sec_one_loop}, we will
be allowed to derive the RG equations at two-loop level. All the
related coefficients are demonstrated in the \ref{Appendix_coefficiets}:
(i) $m_1$, $m_2$, and $m_3$ collect the contribution from Fig.~\ref{Fig_fermion_two_loop} (a);
(ii) $d_1$, $d_2$, and $d_3$ collect the contribution from Fig.~\ref{Fig_fermion_two_loop} (c);
(iii) $f_1$, $f_2$, and $f_3$ collect the contribution from Fig.~\ref{Fig_fermion_two_loop} (e) and
by comparing with results with $C_1$, $C_2$, and $C_3$, we can obtain $f_i=C_i$, ($i=1, 2, 3$);
(iv) $C_h$ collects the contribution from Fig.~\ref{Fig_disorder_two_loop} (g2);
(v) $C_f$ collects the contribution from Fig.~\ref{Fig_disorder_two_loop} (g3);
(vi) $C_k$ collects the contribution from Fig.~\ref{Fig_disorder_two_loop} (i2);
(vii) $C_l$ collects the contribution from Fig.~\ref{Fig_disorder_two_loop} (j4);
(viii) $C_m$ collects the contribution from Fig.~\ref{Fig_disorder_two_loop} (k2);
(ix) $C_n$ collects the contribution from Fig.~\ref{Fig_disorder_two_loop} (k3);
(x) Others Feynman diagrams in Figs. \ref{Fig_fermion_two_loop} and \ref{Fig_disorder_two_loop}
either counteract each other due to the appearance of ladder and crossing diagrams or their
contributions can be expressed by $C_1$, ($i=1,2,3$), $C_g$ and/or above two-loop coefficients.

The Dyson equation for fermion propagator can now be formally
written as
\begin{eqnarray}
G^{-1}_{\psi}(\mathbf{k},\omega)
&=& -i\omega + v_F k_x \tau^z +
v_{\Delta}k_{y}\tau^{x}\nonumber \\
&& - \Sigma^{\mathrm{L1(a)}}(\mathbf{k},\omega)
- \Sigma^{\mathrm{L1(b)}}(\mathbf{k},\omega) \nonumber \\
&& - \Sigma^{\mathrm{L2(a)}}(\mathbf{k},\omega) -
\Sigma^{\mathrm{L2(b)}}(\mathbf{k},\omega)\nonumber \\
&& - \Sigma^{\mathrm{L2(c)}}(\mathbf{k},\omega) -
\Sigma^{\mathrm{L2(d)}}(\mathbf{k},\omega) \nonumber\\
&&-\Sigma^{\mathrm{L2(e)}}(\mathbf{k},\omega) -
\Sigma^{\mathrm{L2(f)}}(\mathbf{k},\omega)\nonumber \\
&& - \Sigma^{\mathrm{L2(g)}}(\mathbf{k},\omega).
\end{eqnarray}
Before computing the RG equations, we need to obtain the expression
of $\eta_f$. Integrating over the momentum shell between $b\Lambda$
and $\Lambda$, we obtain
\begin{widetext}
\begin{eqnarray}
&&\int^{b\Lambda}d^{2}\mathbf{k} d\omega \psi^{\dagger}(-i\omega)
\left\{1-(C_1-2C_g)l-\left[\frac{m_1}{6}-C_gd_1+C_g\left(C_1-\frac{C_2}{2}-\frac{C_3}{2}\right)
-2C_gf_1-2C^2_g-2C^2_g\right]l^2\right\}\psi\nonumber\\
&=&\int^{b\Lambda}d^{2}\mathbf{k} d\omega \psi^{\dagger}
(-i\omega)\left\{1-(C_1-2C_g)l-\left[\frac{m_1}{6}-C_gd_1+C_g\left(C_1-\frac{C_2}{2}-\frac{C_3}{2}\right)
-2C_gf_1-4C^2_g\right]l^2\right\}\psi\nonumber\\
&\approx&\int^{b\Lambda}d^{2}\mathbf{k} d\omega\psi^{\dagger}
(-i\omega)\exp\left\{(2C_g-C_1)l-\left[\frac{m_1}{6}-C_gd_1+C_g\left(C_1-\frac{C_2}{2}-\frac{C_3}{2}\right)
-2C_gf_1-4C^2_g\right]l^2\right\}\psi,
\end{eqnarray}

\begin{figure}
\centering
\includegraphics[width=6.0in]{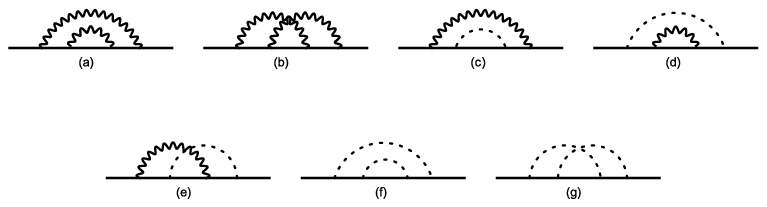}
\vspace{-0.10cm} \caption{Two-loop corrections to the fermion
self-energy in the replica limit $R\rightarrow0$.}\label{Fig_fermion_two_loop}
\end{figure}
\begin{figure}[t]
\centering
\epsfig{file=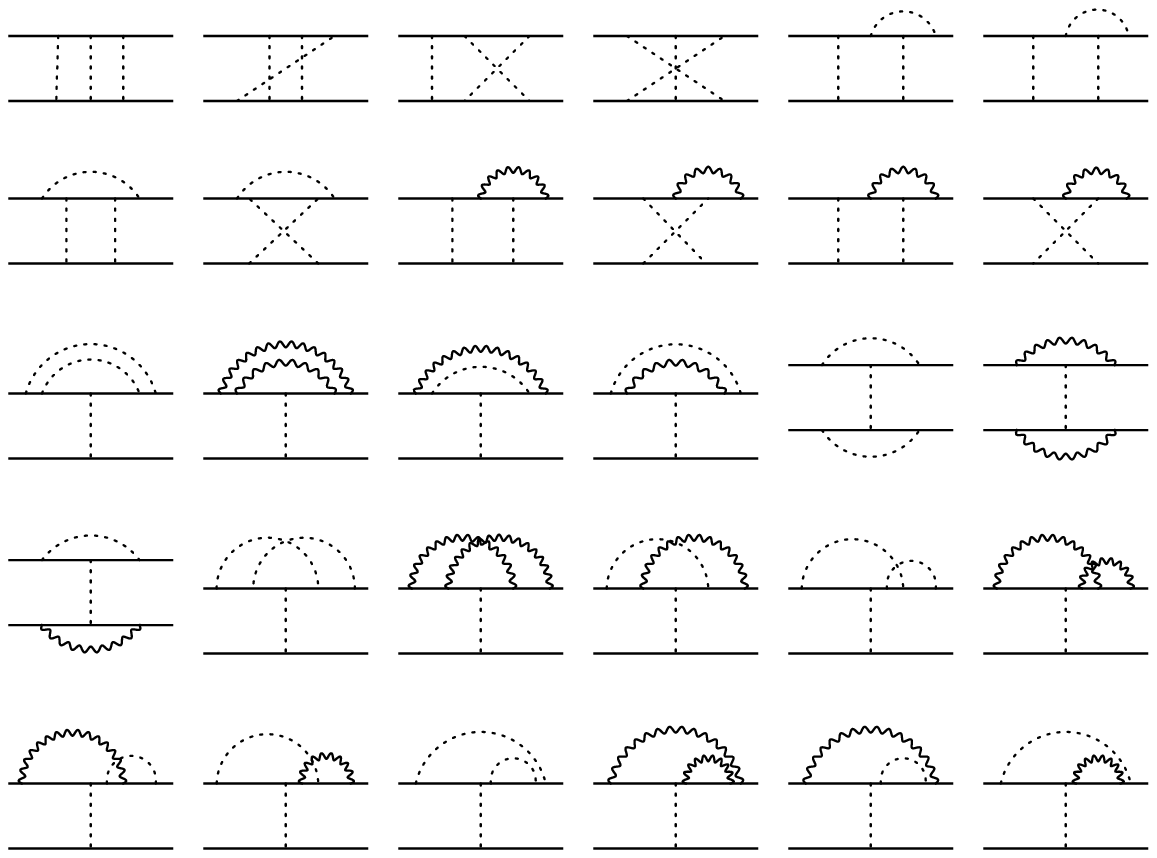,height = 13.18cm,width=16.8cm}
\vspace{-0.05cm}
\caption{Two-loop corrections to the disorder vertex in the replica
limit $R\rightarrow0$.}\label{Fig_disorder_two_loop}
\end{figure}

\end{widetext}
where $m_1$ and $f_1$ come from the contribution of Fig.~\ref{Fig_fermion_two_loop} (a)

After scaling transformation, this term should go back to the free
form, thus
\begin{eqnarray}
\hspace{-0.8cm}&& - \int^l_0 \eta_f dl + \left\{(2C_g-C_1)l -
\left[\frac{m_1}{6} - C_g d_1\right.\right.\nonumber\\
\hspace{-0.8cm}&&\left.\left.+C_g\left(C_1-\frac{C_2}{2}-\frac{C_3}{2}\right) -
2C_g f_1 -4C^2_g\right]l^2\right\} = 0.
\end{eqnarray}
Therefore, $\eta_f$ can be obtained at the two-loop level,
\begin{eqnarray}
\eta_{f}
&=&(2C_{g}-C_{1})-\Bigl[\frac{m_1}{3}+C_g\Bigl(2C_1-C_2-C_3-2d_1\nonumber\\
&&-4f_1-8C_g\Bigr)\Bigr]l.
\end{eqnarray}

\subsection{RG equations of fermion velocities}

After renormalization and rescaling transformation, the Fermi
velocity term $v_F k_x \tau^z$ must return to its original term,
namely
\begin{eqnarray}
&&\int^{b\Lambda}d^{2}\mathbf{k}d\omega\psi^{\dagger}(v_Fk_x\tau^z)\nonumber\\
&&\times\left\{1-C_2l-\left[\frac{m_2}{6} - C_g d_2
- 2C_g f_2\right]l^2\right\}\psi \nonumber\\
&\approx&\int^{b\Lambda}d^{2}\mathbf{k} d\omega\psi^{\dagger}
(v_Fk_x\tau^z)\nonumber\\
&&\times\exp\left\{-C_2l-\left[\frac{m_2}{6} - C_g(d_2
+ 2f_2)\right]l^2\right\}\psi \nonumber\\
&=& \int^{\Lambda}d^{2}\mathbf{k}d\omega\psi^{\dagger}(v_F
k_x\tau^z)e^{-4l}e^{4l-\int^{l}_0\eta^{\mathrm{L2}}_fdl} \nonumber\\
&&\times\exp\left\{-C_2l-\left[\frac{m_2}{6}-C_g(d_2+2f_2)
\right]l^2\right\}\psi \nonumber\\
&=& \int^{\Lambda}d^{2}\mathbf{k}d\omega \psi^{\dagger}(v_F
k_x\tau^z)\nonumber\\
&&\times\exp\left\{-(2C_g-C_1)l + \left[\frac{m_1}{6} - C_g
d_1\right.\right.\nonumber\\
&&\left.\left.+C_g\left(C_1-\frac{C_2}{2}-\frac{C_3}{2}\right) -2C_g
f_1 - 4C^2_g\right]l^2\right\} \nonumber\\
&&\times \exp\left\{-C_2l-\left[\frac{m_2}{6}-C_g(d_2+2f_2)
\right]l^2\right\}\psi \nonumber \\
&\equiv& \int^{\Lambda}d^{2}\mathbf{k}d\omega \psi^{\dagger}(v'_F
k_x \tau^z)\psi,
\end{eqnarray}
where the renormalized Fermi velocity $v'_F$ becomes scale-dependent
and is determined by the following expression
\begin{widetext}
\begin{eqnarray}
v'_F &=& v_F \exp\left\{(C_1-C_2-2C_g)l + \left[\frac{(m_1-m_2)}{6}
+ C_g\left(C_1 - \frac{C_2}{2}\right.\right.\right.\nonumber\\
&&\left.\left.\left.-\frac{C_3}{2}- d_1 - 2f_1 + (d_2+2f_2) -
4C_g\right)\right]l^2\right\}.
\end{eqnarray}
It is now easy to extract the RG equation for $v_F$ up to two-loop level,
\begin{eqnarray}
\frac{dv_F}{dl} &=& \left\{(C_1-C_2-2C_g) +
\left[\frac{(m_1 - m_2)}{3} + 2C_g \left(2C_2 - C_1 + d_2 -
d_1-4C_g\right) - C_g\left(C_2 + C_3\right)\right]l\right\}v_F,\label{Eq_vF_two_loop}
\end{eqnarray}
where the identity $f_i=C_i$ is used. The two-loop RG equation for the gap
velocity $v_{\Delta}$ can be analogously obtained, i.e.,
\begin{eqnarray}
\frac{dv_{\Delta}}{dl} &=& \left\{(C_1-C_3-2C_g) +
\left[\frac{(m_1-m_3)}{3} + 2C_g\left(2C_3-C_1+d_3-d_1-4C_g\right) - C_g(C_2+C_3)\right]l\right\}v_{\Delta}.\label{Eq_vD_two_loop}
\end{eqnarray}
Based on the above two equations, we can also derive the RG equation
for the velocity ratio up to two-loop level,
\begin{eqnarray}
\hspace{-0.99cm}\frac{d\frac{v_{\Delta}}{v_F}}{dl} =
\left\{(C_2-C_3)+\left[\frac{(m_2-m_3)}{3} + 2C_g\left(2C_3-2C_2 + d_3 - d_2\right)\right]l\right\}
\frac{v_{\Delta}}{v_F}.\label{Eq_vDF_two_loop}
\end{eqnarray}
\begin{figure}
\centering
\includegraphics[width=3.6in]{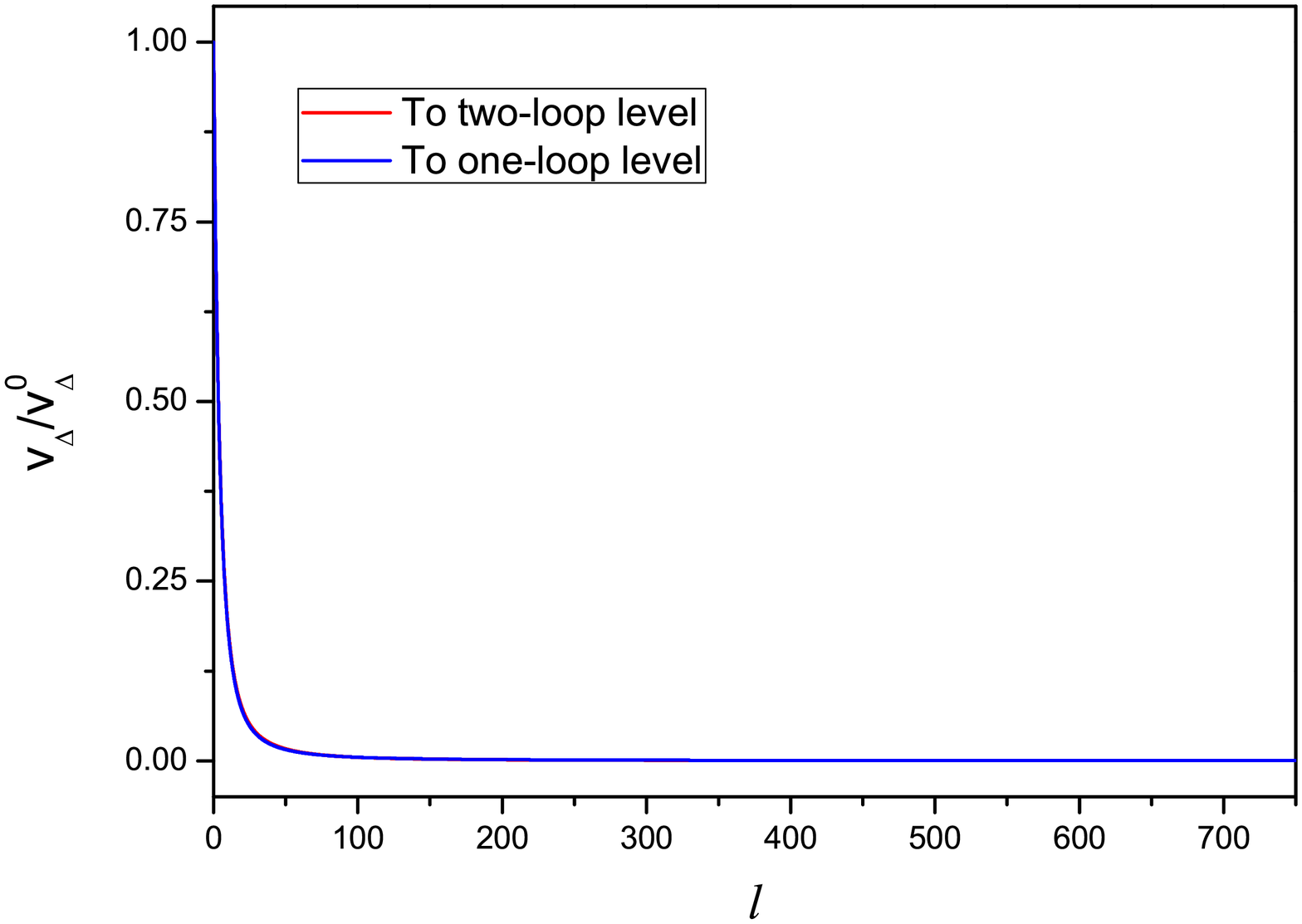}\hspace{-0.5cm}
\includegraphics[width=3.6in]{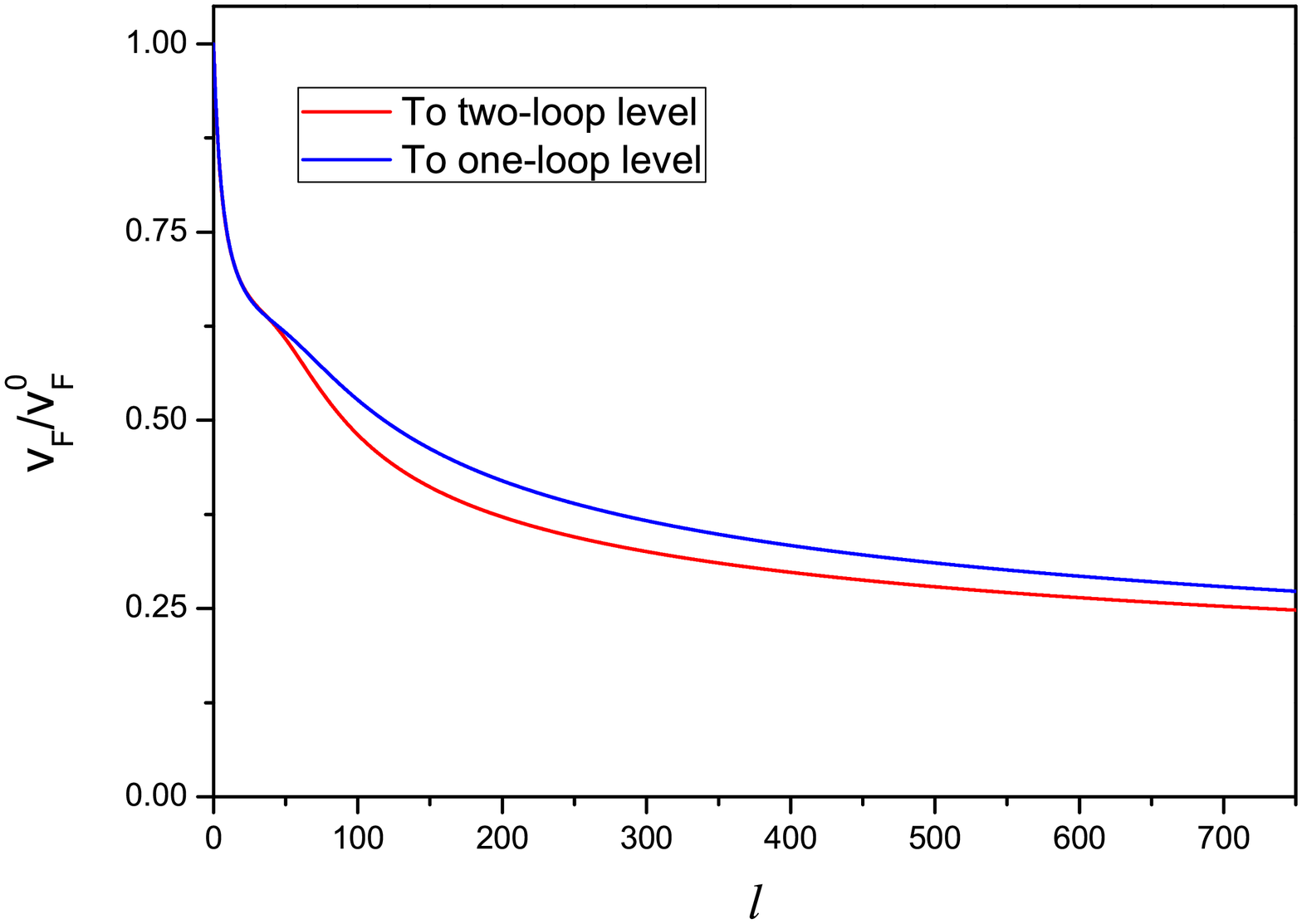}
\vspace{-0.50cm} \caption{The running of $v_\Delta$ (left) and $v_F$
(right) without disorder at the two-loop
level.}\label{Fig_vDvF_two_loop_level_without_disorder}
\end{figure}

\end{widetext}

\begin{figure}
\centering
\includegraphics[width=3.6in]{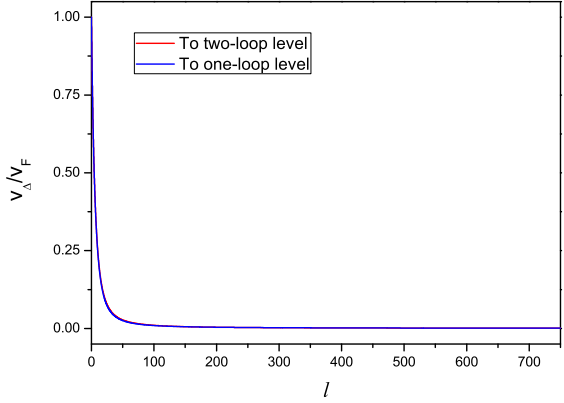}
\vspace{-0.80cm}
\caption{The running of $v_\Delta/v_F$ without disorder at the two-loop
level.}\label{Fig_vD_vF_two_loop_level_without_disorder}
\end{figure}

\subsection{RG equation of disorder strength parameter $\zeta$}

The strength parameter $v_\Gamma$ of random chemical potential
enters into the above equations for fermion velocities. Due to the
interplay of nematic fluctuation and disorder scattering, this
parameter also flows under RG transformation. We calculate the
two-loop diagrams for the correction to fermion-disorder vertex
function and combine them with one-loop contribution, and then
obtain

\begin{eqnarray}
&&\int^{b\Lambda}d^2\mathbf{k} d^2\mathbf{k'}d^2\mathbf{k''}d\omega
d\omega'' \psi^{\dagger}_m(\mathbf{k},\omega)
\psi_m(\mathbf{k'},\omega)\psi^{\dagger}_n(\mathbf{k''},\omega'')
\nonumber \\
&&\times \psi_n(\mathbf{k}+\mathbf{k''}-\mathbf{k'},\omega'')
\left(-\frac{\zeta}{4}\right)\Bigl[1-2C_{1}l+4C_{g}l+4C_g^2l^2\nonumber \\
&&+C_ll^2 +C_1 C_h l^2 + C_g C_f l^2 + C_k l^2 -2C_g C_1 l^2
+ 8C^2_g l^2\nonumber \\
&& + 2C^2_1 l^2 + C_m l^2 - 4C_1 C_g l^2 - 8C_g^2 l^2 + 16C_g^2 l^2 + C_g C_n l^2\nonumber\\
&& + 2C_g(C_2+C_3)l^2\Bigr] \nonumber \\
&=&\int^{\Lambda}d^2\mathbf{k}d^2\mathbf{k'}d^2\mathbf{k''}d\omega
d\omega''\psi^{\dagger}_m(\mathbf{k},\omega)\psi_m(\mathbf{k'},\omega)
\psi^{\dagger}_n(\mathbf{k''},\omega'')\nonumber\\
&&\times\psi_n(\mathbf{k}+\mathbf{k''}-\mathbf{k'},\omega'')
\left(-\frac{\zeta}{4}\right)\exp\Bigl\{\Bigl[\frac{m_1}{3} - 2C_g
d_1\nonumber\\
&& + 12C_g^2 + C_1 C_h  + C_g C_f + 2C^2_1 - 8C_1 C_g + C_k + C_l\nonumber\\
&& + C_m + C_g C_n + C_g(C_2 + C_3)\Bigr]l^2\Bigr\}\nonumber \\
&\equiv& \int^{\Lambda}d^2\mathbf{k} d^2\mathbf{k'} d^2\mathbf{k''}
d\omega d\omega'' \psi^{\dagger}_m(\mathbf{k},\omega)
\psi_m(\mathbf{k'},\omega)\psi^{\dagger}_n(\mathbf{k''},\omega'')\nonumber\\
&&\times \psi_n(\mathbf{k}+\mathbf{k''}-\mathbf{k'},\omega'')
\left(-\frac{\zeta'}{4}\right),
\end{eqnarray}
where
\begin{eqnarray}
\zeta' &=& \zeta\exp\Bigl\{\Bigl[\frac{m_1}{3} -2C_g d_1 + 12C_g^2
+ C_1 C_h + C_g C_f\nonumber\\
&& + 2C^2_1 - 8C_1 C_g  + C_k + C_l + C_m + C_g C_n\nonumber\\
&& + C_g(C_2 + C_3)\Bigr]l^2\Bigr\}.
\end{eqnarray}
The RG equation for parameter $\zeta$ up to two-loop level is
finally found to be
\begin{widetext}
\begin{eqnarray}
\frac{d\zeta}{dl} &=& 2l\Bigl[C_g(12C_g - 2d_1 + C_f) + C_1(C_h +
2C_1 - 8C_g) +\left(\frac{m_1}{3} + C_k + C_l + C_m\right) + C_g(C_n + C_2 +
C_3)\Bigr]\zeta.\label{Eq_disorder_two_loop}
\end{eqnarray}

\begin{figure}
\centering
\includegraphics[width=3.6in]{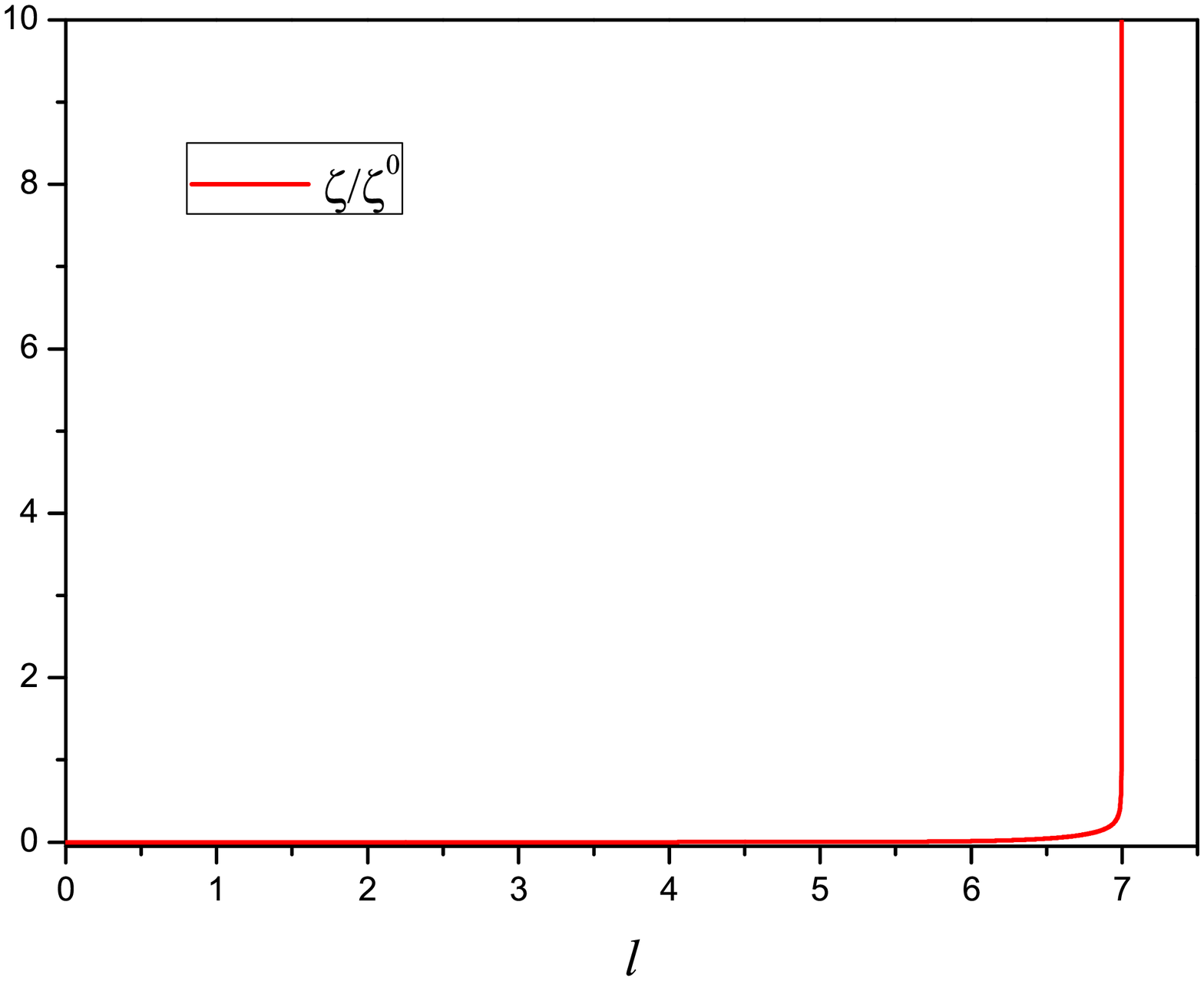}\hspace{-0.8cm}
\includegraphics[width=3.6in]{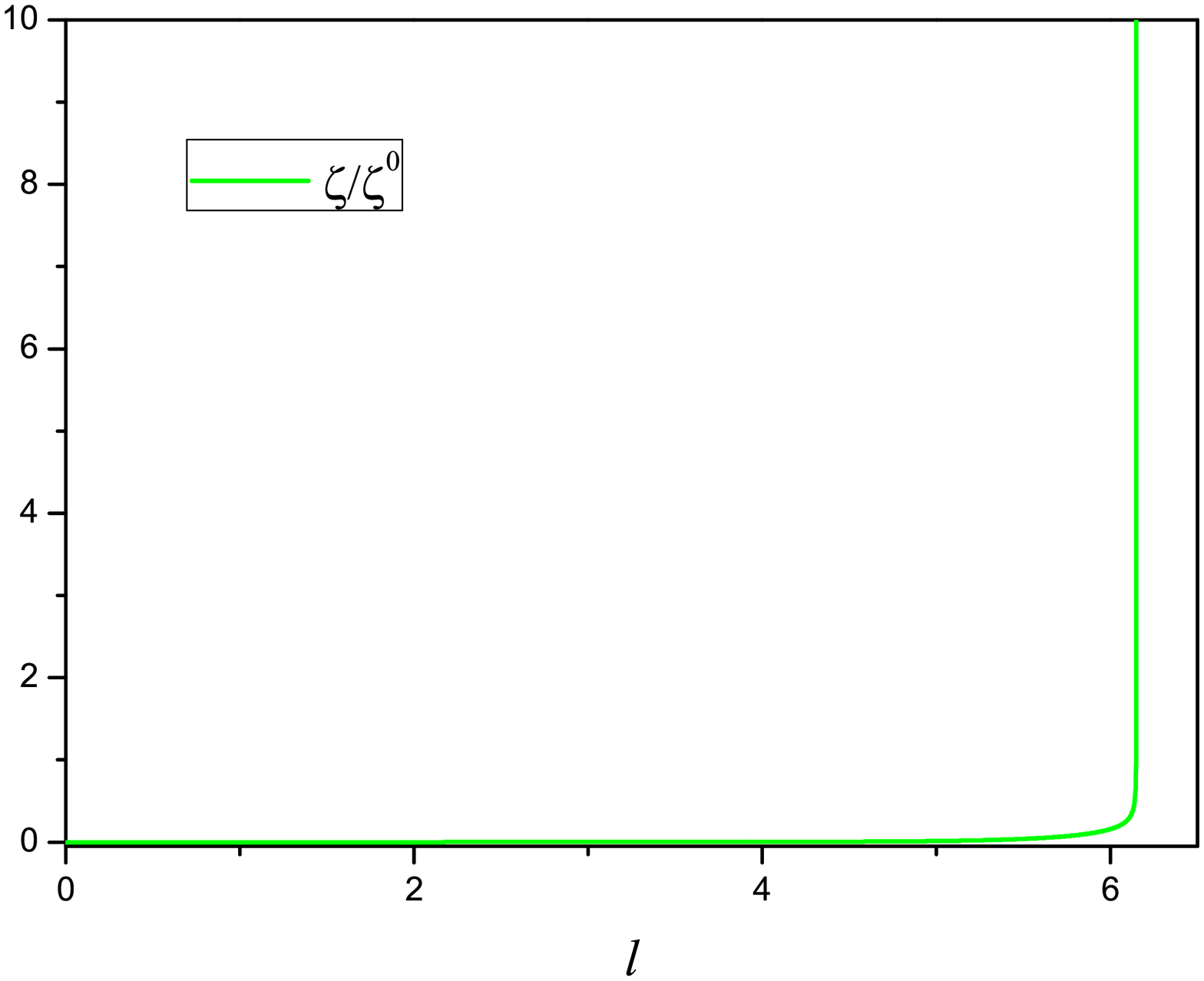}
\vspace{-0.50cm} \caption{Flows of the disorder strength $\zeta$ with
the initial values $(v_\Delta/v_F)^0=0.1,v^0_F=1,v^0_\Delta=0.1$,
$\zeta^0=10^{-5}$ (left) and $\zeta^0=10^{-4}$ (right) at the
two-loop level.} \label{Fig_Two_loop_with_disorder_mE5}
\end{figure}

\begin{figure}
\centering
\includegraphics[width=3.6in]{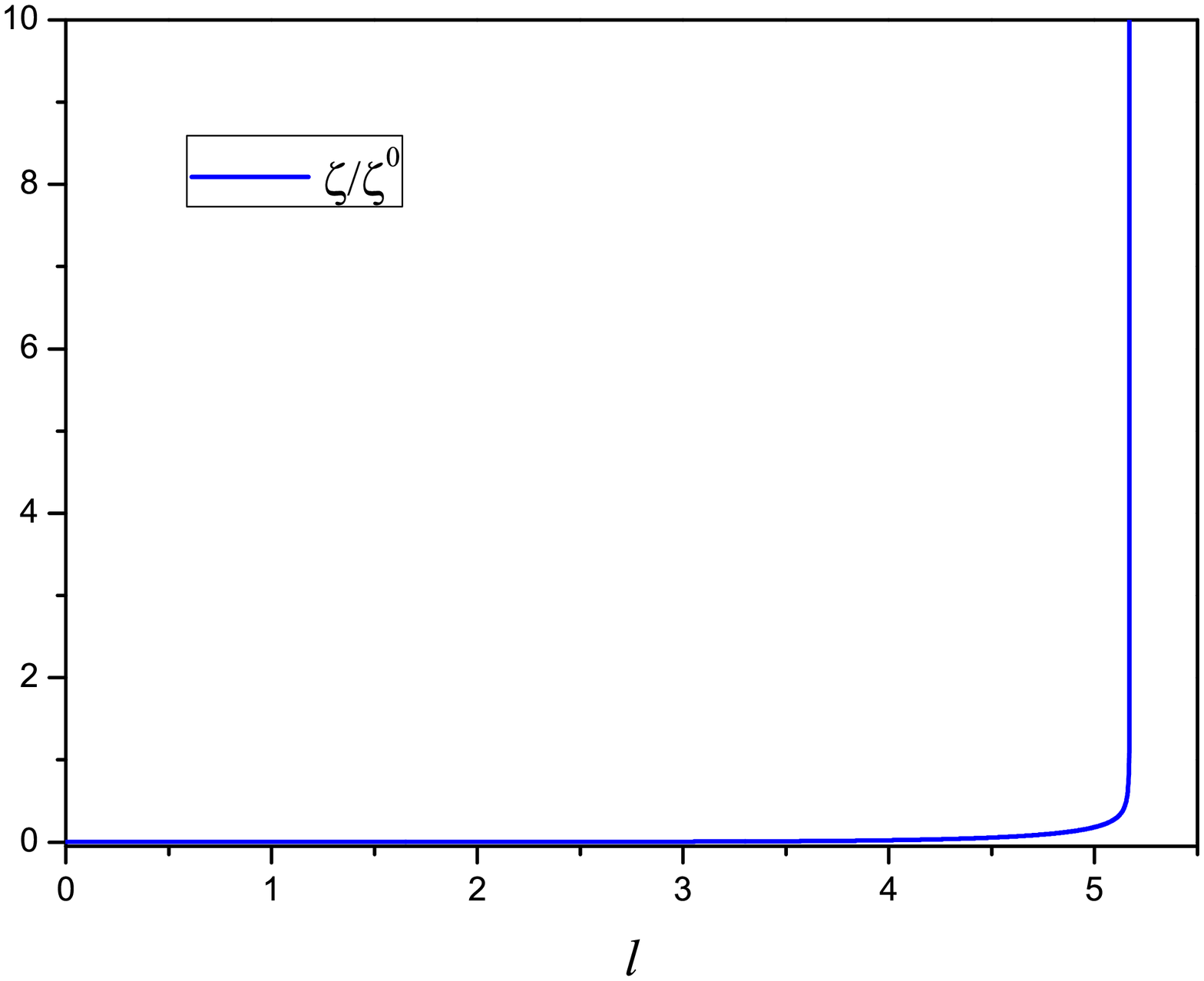}\hspace{-0.8cm}
\includegraphics[width=3.6in]{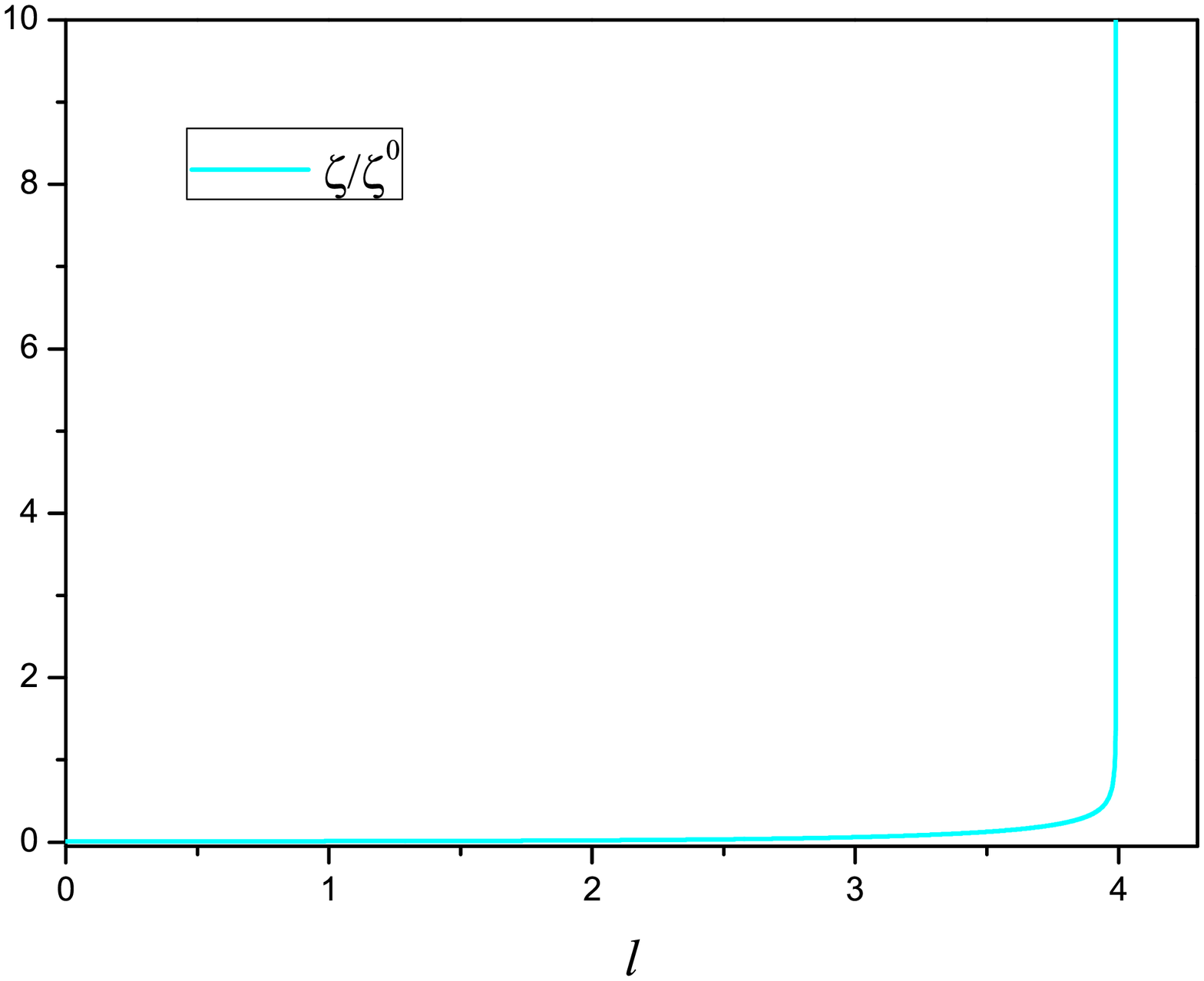}
\vspace{-0.5cm} \caption{Flows of $\zeta$ with initial values
$(v_\Delta/v_F)^0=0.1,v^0_F=1,v^0_\Delta=0.1$, $\zeta^0=10^{-3}$
(left) and $\zeta^0=10^{-2}$ (right) at the two-loop
level.}\label{Fig_Two_loop_with_disorder_mE3}
\end{figure}

\end{widetext}

\section{Numerical solutions for RG equations and discussions}
\label{Sec_numerical_discussions}

The RG equations for fermion velocities and disorder parameter have
been derived in the last section. The aim of this section is to
analyze the solutions of these equations, which are
self-consistently coupled to each other and therefore need to be
solved numerically. To make the discussion more transparent, we
first consider the clean limit with $\zeta = 0$ and then include
disorder later.

\subsection{Clean limit}

In the clean case, $\zeta = 0$, the RG equations have the following
form up to two-loop level,
\begin{eqnarray}
\frac{dv_F}{dl} &=& \left[C_1-C_2 +
\frac{(m_1-m_2)}{3}l\right]v_F,\\
\frac{dv_{\Delta}}{dl}
&=&\left[C_1-C_3+\frac{(m_1-m_3)}{3}l\right]v_{\Delta},\\
\frac{d\frac{v_{\Delta}}{v_F}}{dl} &=& \left[C_2-C_3 +
\frac{(m_2-m_3)}{3}l\right]\frac{v_{\Delta}}{v_F}.
\end{eqnarray}
Solving these RG equations numerically, we find that the results
qualitatively agrees with those obtained at one-loop level
Ref.~\cite{Huh2008PRB}, which can be clearly seen from
Figs.~(\ref{Fig_vDvF_two_loop_level_without_disorder}) and
(\ref{Fig_vD_vF_two_loop_level_without_disorder}). In particular,
$v_{F}$ vanishes with growing $l$ much more slowly than
$v_{\Delta}$, so the velocity ratio $v_{\Delta}/v_{F} \rightarrow 0$
at large length scale, which corresponding to a stable fixed point
of extreme anisotropy. It therefore turns out that the one-loop
results are robust and that the $1/N$-expansion is fairly reliable.

\subsection{In the presence of disorders}

Now we consider the influence of random chemical potential. Before
running into complex numerical computation, it is helpful to first
make a simple analysis about the behavior of parameter $\zeta$. From
the flow equation of $\zeta$ (\ref{Eq_disorder_two_loop}), it is
convenient to extract a general differential equation,
$\frac{d\zeta}{dl} = Cl\zeta$. Apparently, the fate of $\zeta$ in
the low-energy region (i.e., large $l$ limit) is unambiguously
determined by the value of the coefficient $C$: as the running
length scale $l$ increases, $\zeta$ flows towards infinity,
vanishes, and stays at a certain constant for $C>0$, $C<0$, and
$C=0$, respectively. Previous one-loop calculations
\cite{Stauber2005PRB, WLK2011PRB} showed that the coefficient $C$
vanishes, so the strength parameter $\zeta$ of random chemical
potential is marginal and remains a constant as $l$ varies. After
including two-loop corrections, $C$ may become either positive or
negative, which will be addressed by solving the coupled flow
equations (\ref{Eq_vF_two_loop}), (\ref{Eq_vD_two_loop}),
(\ref{Eq_vDF_two_loop}), and (\ref{Eq_disorder_two_loop}).

The numerical solutions of RG equations with properly chosen initial
values of disorder parameter are shown in
Figs.~\ref{Fig_Two_loop_with_disorder_mE5} and
\ref{Fig_Two_loop_with_disorder_mE3}. As $l$ grows, $\zeta$
increases quickly and eventually diverges in the limit $l\rightarrow
+\infty$. This behavior does not depend on the initial values of the
running parameters. Therefore, the strength parameter of random
chemical potential, which is marginal at one-loop level, becomes
relevant after including the two-loop order corrections. This change
may have significant influence on the physical properties of the
system under consideration since even an infinitesimal relevant
parameter will flow to very large at ultra low energies.

A natural question arises: how can we understand the marginally
relevant nature of the characteristic parameter $\zeta$ of random
chemical potential and its impacts on the nematic quantum
criticality in $d$-wave superconductors? In the clean limit, $\zeta
= 0$, the strong quantum fluctuation of nematic order results in
non-Fermi liquid behavior of nodal fermions \cite{Kim2008PRB,
Liu2015}. Once random chemical potential is introduced, the
non-Fermi liquid state is immediately turned into a diffusive
metallic state \cite{Fradkin1986PRB, Lee1993PRL, Foster2008PRB,
Shindou2009PRB, Moon2014, Ominato2014PRB, Goswami2011PRL, Kobayashi2014PRL,
Sbierski2014PRL, Lai_Roy2014}, which has a finite zero-energy
density of states and finite scattering rate. This conclusion is
valid for all possible values of $\zeta$, since $\zeta$,
irrespective of its initial value, unavoidably flows to infinity in
the low-energy regime.

\section{Physical implications}\label{Sec_diagram implications}

In addition to the tendency of non-Fermi liquid behavior due to the marginally relevant behavior of random
chemical potential, we, within this section, try to delineate investigate the influence of nematic fluctuation, gapless QPs and marginally relevant random chemical potential on a number of significantly physical observables in a $d$-wave cuprate superconductor at the nematic QCP, such as the superfluid density, critical temperature, and thermal conductivity.

Learning from the Fig.~\ref{Fig_Two_loop_with_disorder_mE5} and the discussion in Sec.~\ref{Sec_numerical_discussions}, we can be recalled some points. On the one hand,
the strength of random chemical potential, due to the marginally relevant behavior, flows away
with the energy scale $l$ reaching a critical value, dubbed $l^*$, at the nematic QCP. This
indirectly triggers the unphyscial behaviors of fermion velocities $v_F$ and $v_\Delta$ at a
very scale $l=l^*$ which is dependent on the initial values of $v_F$ and $v_\Delta$ by the
growing strength of random chemical potential and will be carefully investigated in the following
subsections.  On the other hand, runaway behavior of physical quantities suggests a first-order phase transition~\cite{Domany1977PRB,Chen1978PRB,Rudnick1978PRB,Iacobson1981AP,Cardybook}. In order to
figuratively describe it in finite temperatures, we employ a useful formula $T=T_ce^{-l}$ \cite{Huh2008PRB,She2014} ($T_c$ is the critical temperature of superconductor) to translate
$l^*$ to $T^*$.

It is necessary to further emphasize that our effective theory and method are only valid for the
continuous phase transitions and hence we subsequently focus on the critical behaviors of physical
observables with the confined region $T>T^*$, such as superfluid density, critical temperature and
thermal conductivity.

\subsection{Superfluid density and critical temperature}

\begin{figure}[t]
\centering
\epsfig{file=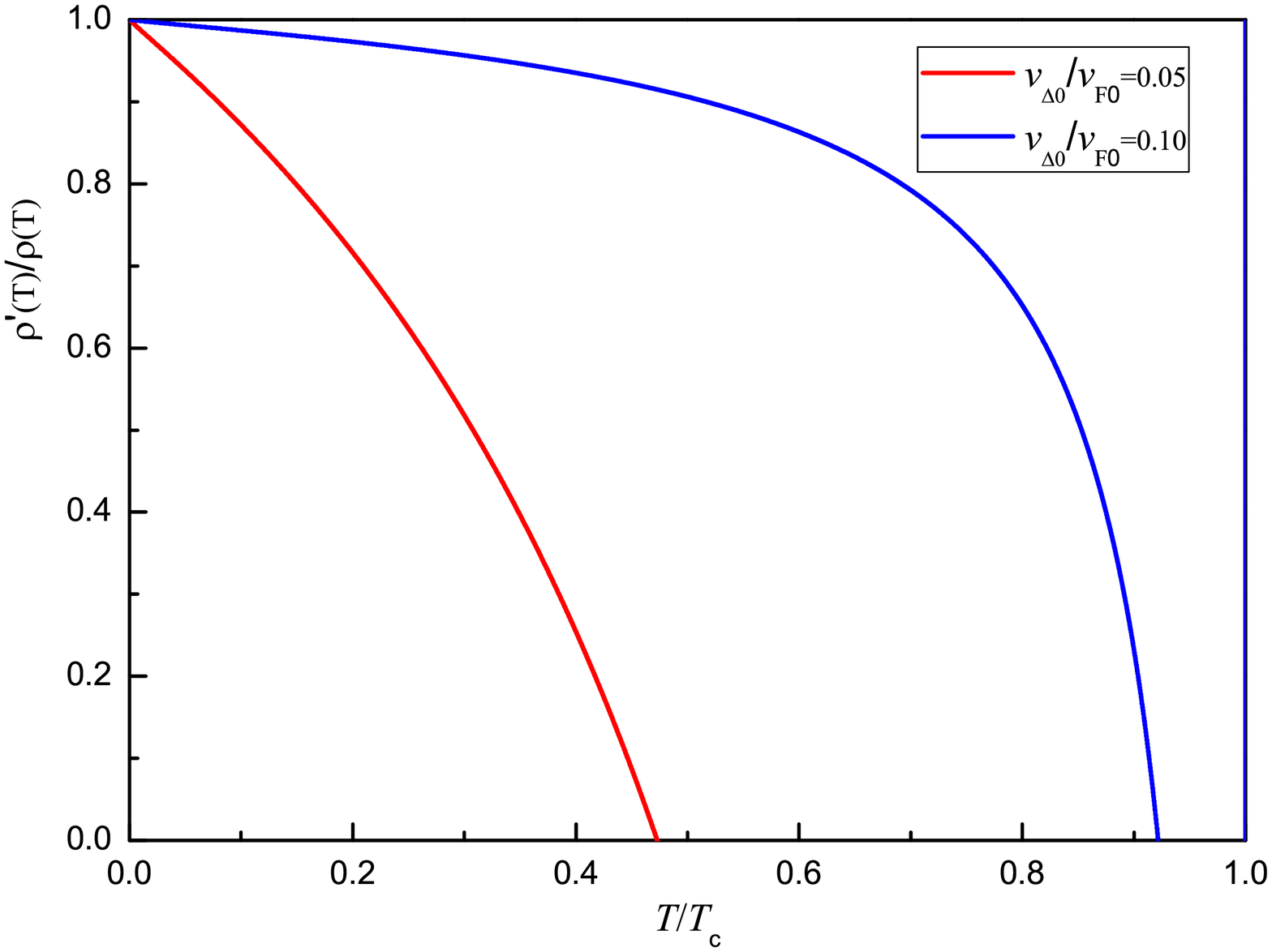,height = 7.05cm,width=9.8cm}
\vspace{-0.90cm}
\caption{Temperature-dependence superfluid density in the presence of the nematic fluctuation and random
chemical potential at nematic QCP for the representatively initial values of $v_\Delta/v_F$ and random
chemical potential. The primary conclusions are independent of the initial values.}\label{Fig_density}
\end{figure}

As studied by Lee and Wen~\cite{Lee1997PRL}, the superfluid density in the noninteracting case
exhibits a linear temperature dependence which is in agreement with experiments~\cite{Hardy1993PRL}
\begin{eqnarray}
\frac{\rho_s(T)}{m}=\frac{\rho_s(0)}{m}-\frac{2\ln 2}{\pi}\frac{v_F}{v_\Delta}T,
\end{eqnarray}
with $\rho_s(0)=\frac{x}{a^2}$ where $x$ and $a$ represent the doping concentration and lattice spacing, respectively and $T_c\propto\frac{v_\Delta}{v_F}\frac{x}{ma^2}$ by defining $\rho_s(T_c)=0$, reproducing
the Uemura plot~\cite{Uemura1989PRL}. The other term of right hand side is the contribution of the normal
QPs density. As presented in the previous section, the fermion velocities $v_F$, $v_\Delta$, and the ratio $v_\Delta/v_F$ are heavily renormalized by the interplay between nematic fluctuation and the effects of
random chemical potential in the vicinity of the nematic QCP. We can approximately derive the superfluid
density of of the $d$-wave superconducting state at the nematic QCP which is renormalized by nematic
fluctuation and marginally relevant random chemical potential~\cite{Lee1997PRL,Durst2000PRB},
\begin{eqnarray}
\rho'_s(T)=\rho_s(0)-\rho_n(T),\label{Eq_rho}
\end{eqnarray}
where the normal QPs density is complicatedly dependent on the fermion velocities $v_F$ and $v_\Delta$
\begin{eqnarray}
\frac{\rho_n(T)}{m}=\frac{4}{k_BT}\int \frac{d^2\mathbf{k}}{(2\pi)^2}
\frac{v_F^2(\mathbf{k})e^{\frac{\sqrt{v_F^2(\mathbf{k})k_x^2+v_\Delta^2(\mathbf{k})k_y^2}}{k_BT}}}
{\left(1+e^{\frac{\sqrt{v_F^2(\mathbf{k})k_x^2+v_\Delta^2(\mathbf{k})k_y^2}}{k_BT}}\right)^2}.\label{Eq_rho_n}
\end{eqnarray}
By combing the Eqs. (\ref{Eq_rho}) and (\ref{Eq_rho_n}) and Eqs. (\ref{Eq_vF_two_loop}), (\ref{Eq_vD_two_loop}), and (\ref{Eq_vDF_two_loop}), the superfluid density and critical temperature are dependent on the running of velocities $v_F$ and $v_\Delta$ and also the behavior of disorder which is coupled with the velocities $v_F$
and $v_\Delta$ as lineated by Eqs. (\ref{Eq_vF_two_loop}), (\ref{Eq_vD_two_loop}), and (\ref{Eq_vDF_two_loop}). To include these corrections and after some numerical calculations, we obtain the $\rho'_s(T)/\rho_s(T)$ with
the different initial values of fermion velocities $v_F$ and $v_\Delta$ for $T>T^*$ as depicted in Fig.~\ref{Fig_density}.

Studying from the numerical results in Fig.~\ref{Fig_density}, we can draw a conclusion that
the superfluid density and critical temperature are largely suppressed in the presence of
marginally relevant random chemical potential nearby the nematic QCP. This may be very
instructive to provide another clue to locate the nematic QCP.

\subsection{Thermal conductivity}

\begin{figure}[t]
\centering
\epsfig{file=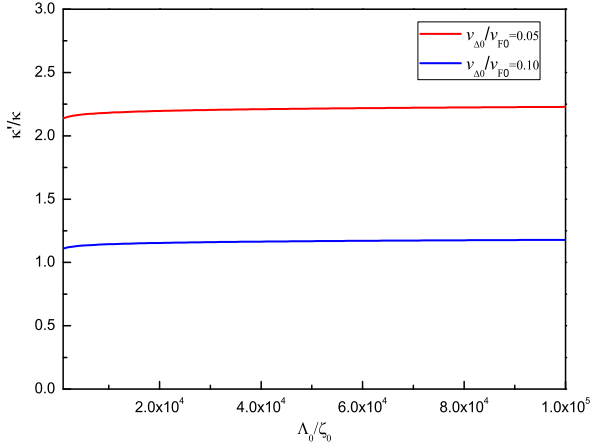,height = 7.05cm,width=9.8cm}
\vspace{-0.90cm} \caption{Thermal conductivity of QPS in the presence of nematic fluctuation and
random chemical potential at nematic QCP for the representatively initial values of $v_\Delta/v_F$
and random chemical potential. The primary conclusions are independent of the initial values.}\label{Fig_thermal}
\end{figure}

We then turn to the thermal conductivity of QPS. By paralleling the derivations in Ref.~\cite{Durst2000PRB},
we obtain the thermal conductivity of QPs for the constant values of fermion velocities $v_F$ and $v_\Delta$,
\begin{eqnarray}
\frac{\kappa}{T}&=&\frac{k^2_B}{3}
\left(\frac{v_f}{v_\Delta}+\frac{v_\Delta}{v_f}\right),\label{Eq_kappa}
\end{eqnarray}
and for the energy scale-dependence $v_F$ and $v_\Delta$ which are comprised the nematic fluctuation
and marginally relevant random chemical potential in the vicinity of nematic QCP,
\begin{eqnarray}
\frac{\kappa'}{T}
&=&\left(\frac{k^2_B}{3}\right)\int^{\frac{\Lambda_0}{\zeta_0}}_0 \!\!\frac{d\mathbf{k}}
{\left(1+\mathbf{k}^2\right)^2}\left[\frac{v_F(\mathbf{k})}
{v_\Delta(\mathbf{k})}+\frac{v_\Delta(\mathbf{k})}
{v_F(\mathbf{k})}\right]\!,\label{Eq_kappa_prime}
\end{eqnarray}
with $\Lambda_0$ determined by the lattice constant and $\zeta_0$ the initial value of disorder strength.
In order to examine the influence of nematic fluctuation and random chemical potential, we numerically
carry out the Eqs. (\ref{Eq_vF_two_loop}), (\ref{Eq_vD_two_loop}), (\ref{Eq_vDF_two_loop}), (\ref{Eq_kappa})
and (\ref{Eq_kappa_prime}) and reach the numerical results as presented for $T>T^*$ in Fig.~\ref{Fig_thermal}.

As exhibited in Fig.~\ref{Fig_thermal}, the thermal conductivity of QPs, influenced by nematic fluctuation
and marginally relevant random chemical potential at the nematic QCP, is highly increased and insensitive
to the initial values of random chemical potential. This may be helpful to enhance our understand the phase diagram of $d$-wave superconductors.

\section{Summary}\label{Sec_summary}

In summary, we have investigated the impact of a particular
disorder, namely random chemical potential, on the stability of
nematic quantum critical point located in the superconducting dome
of $d$-wave superconductors. We performed a detailed RG analysis up
to two-loop order within an effective field theory that describes
the interplay of nematic fluctuation and disorder scattering. The
effective parameter that characterize the strength of random
chemical potential is marginal at one-loop order, but becomes
relevant due to the inclusion of two-loop order corrections. This
means that, even if we start from a very weak disorder, the
effective strength of disorder eventually becomes infinitely large
at the lowest energy. The Dirac fermions therefore enter into a
diffusive metallic state driven by random chemical potential at the
nematic quantum critical point \cite{Fradkin1986PRB, Shindou2009PRB,
Ominato2014PRB, Goswami2011PRL, Kobayashi2014PRL, Sbierski2014PRL}.
Furthermore, we carefully study the critical behaviors for a number of
significantly physical observables in a $d$-wave cuprate superconductor
at the nematic QCP, such as the superfluid density, critical temperature,
and thermal conductivity in the vicinity of nematic QCP.

\section*{ACKNOWLEDGEMENTS}

I am very grateful to Guo-Zhu Liu for correlated collaborations and
useful discussions. This work is supported by the China Postdoctoral
Science Foundation under Grant No. 2014M560510, the Fundamental Research
Funds for the Central Universities (P.R. China) under Grant No. WK2030040074
and the National Science Foundation of China under Grant No. 11274286.

\appendix

\section{Some coefficients used in the main text}\label{Appendix_coefficiets}

\begin{widetext}
The coefficients for one-loop Feynman diagrams are
\begin{eqnarray}
C_1&=&\frac{2v_\Delta}{N_f\pi^3v_F}\int^{\infty}_{-\infty}dx
\int^{2\pi}_{0} d \theta\frac{x^2-\cos^2\theta-\left(\frac{v_\Delta}{v_F}\right)^2
\sin^2\theta}{\left[x^2+\cos^2\theta+\left(\frac{v_\Delta}{v_F}\right)^2
\sin^2\theta\right]^2}\mathcal {G}(x,\theta),\\
C_2&=&\frac{2v_\Delta}{N_f\pi^3v_F}\int^{\infty}_{-\infty}dx
\int^{2\pi}_{0}d\theta\frac{\cos^2\theta-x^2-\left(\frac{v_\Delta}{v_F}\right)^2
\sin^2\theta}{\left[x^2+\cos^2\theta+\left(\frac{v_\Delta}{v_F}\right)^2
\sin^2\theta\right]^2}\mathcal{G}(x,\theta),\\
C_3&=&\frac{2v_\Delta}{N_f\pi^3v_F}\int^{\infty}_{-\infty}dx
\int^{2\pi}_{0}d\theta\frac{x^2+\cos^2\theta-\left(\frac{v_\Delta}{v_F}\right)^2
\sin^2\theta}{\left[x^2+\cos^2\theta+\left(\frac{v_\Delta}{v_F}\right)^2
\sin^2\theta\right]^2}\mathcal{G}(x,\theta),\\
\mathcal{G}^{-1}&=&\frac{x^2+\cos^2\theta}{\sqrt
{x^2+\cos^2\theta+\left(\frac{v_\Delta}{v_F}\right)^2\sin^2\theta}}
+\frac{x^2+\sin^2\theta}{\sqrt{x^2+\sin^2\theta
+\left(\frac{v_\Delta}{v_F}\right)^2\cos^2\theta}},
\end{eqnarray}
and
\begin{eqnarray}
C_g=-\frac{\zeta}{4}\frac{1}{2\pi v_{F}v_{\Delta}}.
\end{eqnarray}

For two-loop Feynman diagrams, the corresponding coefficients are
\begin{eqnarray}
m_1&=&\frac{2v_\Delta}{N_f\pi^3v_F}\int^{\infty}_{-\infty}
dx\int^{2\pi}_{0}d\theta\frac{\mathcal{G}(x,\theta)}{\left[x^2+\cos^2\theta
+\left(\frac{v_\Delta}{v_F}\right)^2\sin^2\theta\right]^3}\nonumber\\
&&\times\left\{4x^2\left[C_1x^2+(2C_2-C_1)\cos^2\theta+(2C_3-C_1)\left(\frac{v_\Delta}{v_F}\right)^2
\sin^2\theta\right.\right]\nonumber\\
&&-\left[3C_1x^2+(2C_2-C_1)\cos^2\theta+(2C_3-C_1)\left(\frac{v_\Delta}{v_F}\right)^2
\sin^2\theta\right]\nonumber\\
&&\left.\times\left[x^2+\cos^2\theta+\left(\frac{v_\Delta}{v_F}\right)^2\sin^2\theta\right]\right\},\\
m_2&=&\frac{2v_\Delta}{N_f\pi^3v_F}\int^{\infty}_{-\infty}dx
\int^{2\pi}_{0}d\theta\frac{\mathcal{G}(x,\theta)}{\left[x^2+\cos^2\theta+
\left(\frac{v_\Delta}{v_F}\right)^2\sin^2\theta\right]^3}\nonumber\\
&&\times\left\{\left[(C_2-2C_1)x^2+(C_2-2C_3)\left(\frac{v_\Delta}{v_F}\right)^2\sin^2\theta
-3C_2\cos^2\theta\right]\right.\nonumber\\
&&\times\left[x^2+\cos^2\theta+\left(\frac{v_\Delta}{v_F}\right)^2\sin^2\theta\right]
-4\cos^2\theta\Bigl[(C_2-2C_1)x^2\nonumber\\
&&\left.+(C_2-2C_3)\left(\frac{v_\Delta}{v_F}\right)^2\sin^2\theta-C_2\cos^2\theta\Bigr]\right\},\\
m_3&=&\frac{2v_\Delta}{N_f\pi^3v_F}\int^{\infty}_{-\infty}dx
\int^{2\pi}_{0}d\theta\frac{\mathcal{G}(x,\theta)}{\left[x^2+\cos^2\theta
+\left(\frac{v_\Delta}{v_F}\right)^2\sin^2\theta\right]^3}\nonumber\\
&&\times\left\{\left[3C_3\left(\frac{v_\Delta}{v_F}\right)^2\sin^2\theta+
(2C_2+C_3)\cos^2\theta+(2C_1-C_3)x^2\right]\right.\nonumber\\
&&\times\left[x^2+\cos^2\theta+\left(\frac{v_\Delta}{v_F}\right)^2\sin^2\theta\right]
-4\left(\frac{v_\Delta}{v_F}\right)^2\sin^2\theta\Bigl[(2C_1-C_3)x^2\nonumber\\
&&\left.+C_3\left(\frac{v_\Delta}{v_F}\right)^2\sin^2\theta
+(2C_2+C_3)\cos^2\theta\Bigr]\right\},\\
d_1&=&\frac{2v_\Delta}{N_f\pi^3v_F}\int^{\infty}_{-\infty}dx
\int^{2\pi}_{0}d\theta\frac{\mathcal{G}(x,\theta)}{\left[x^2+\cos^2\theta
+\left(\frac{v_\Delta}{v_F}\right)^2\sin^2\theta\right]^3}\nonumber\\
&&\times\left\{\left[3x^2-\cos^2\theta-\left(\frac{v_\Delta}{v_F}\right)^2
\sin^2\theta\right]\left[x^2+\cos^2\theta+\left(\frac{v_\Delta}
{v_F}\right)^2\sin\theta^2\right]\right.\nonumber\\
&&\left.-4x^2\left[x^2-\cos^2\theta-\left(\frac{v_\Delta}{v_F}\right)^2
\sin^2\theta\right]\right\},\\
d_2&=&\frac{4v_\Delta}{N_f\pi^3v_F}\int^{\infty}_{-\infty}
dx\int^{2\pi}_{0}d\theta\frac{x^2\mathcal{G}(x,\theta)}{\left[x^2+\cos^2\theta
+\left(\frac{v_\Delta}{v_F}\right)^2\sin^2\theta\right]^3}\nonumber\\
&&\times\left[x^2+\left(\frac{v_\Delta}{v_F}\right)^2\sin^2\theta
-3\cos^2\theta\right],\\
d_3&=&-\frac{4v_\Delta}{N_f\pi^3v_F}\int^{\infty}_{-\infty}dx\int^{2\pi}_{0}
d\theta\frac{x^2\mathcal{G}(x,\theta)}{\left[x^2+\cos^2\theta+\left(\frac{v_\Delta}
{v_F}\right)^2\sin^2\theta\right]^3}\nonumber\\
&&\times\left[x^2+\cos^2\theta-3\left(\frac{v_\Delta}{v_F}\right)^2
\sin^2\theta\right],\\
C_h&=&-\frac{2v_\Delta}{3N_f\pi^3v_F}\int^{\infty}_{-\infty}dx\int^{2\pi}_{0}d\theta
\frac{\left(\frac{v_\Delta}{v_F}\right)^2\sin^2\theta\mathcal {G}(x,\theta)}
{\left[x^2+\cos^2\theta+\left(\frac{v_\Delta}{v_F}\right)^2
\sin^2\theta\right]^2},\\
C_f&=&-\frac{4v_\Delta}{N_f\pi^3v_F}\int^{\infty}_{-\infty}dx\int^{2\pi}_{0}d\theta
\frac{\mathcal {G}(x,\theta)}{\left[x^2+\cos^2\theta+\left(\frac{v_\Delta}{v_F}\right)^2
\sin^2\theta\right]^2}\nonumber\\
&&\times\left[x^2-\cos^2\theta-\left(\frac{v_\Delta}{v_F}\right)^2\sin^2\theta\right],\\
C_k&=&\frac{2\sqrt{3\pi}v_{F}v_\Delta}{9N_f^2\pi^3}\int^{\infty}_{-\infty}dx
\int^{2\pi}_{0}d\theta\frac{\mathcal{G}(x,\theta)}{\left[x^2+\cos^2\theta
+\left(\frac{v_\Delta}{v_F}\right)^2\sin^2\theta\right]}\nonumber\\
&&\times\left[E_1x^2-E_2\cos^2\theta+E_3\left(\frac{v_\Delta}
{v_F}\right)^2\sin^2\theta\right],\\
C_l&=&\frac{\zeta}{N_f\pi^4v_F^2}\int^{\infty}_{-\infty}dx\int^{2\pi}_{0}d\theta
\frac{\mathcal{G}(x,\theta)}{\left[x^2+\cos^2\theta+\left(\frac{v_\Delta}
{v_F}\right)^2\sin^2\theta\right]},\\
C_m&=&\frac{4v_\Delta}{3N_f\pi^3v_F}\int^{\infty}_{-\infty}dx\int^{2\pi}_{0}d\theta
\frac{\mathcal{G}(x,\theta)}{\left[x^2+\cos^2\theta+\left(\frac{v_\Delta}{v_F}\right)^2
\sin^2\theta\right]^3}\nonumber\\
&&\times\left\{C_3\left(\frac{v_\Delta}{v_F}\right)^4\sin^4\theta
+3(C_1+C_2)x^2\cos^2\theta-C_2\cos^4\theta-C_1x^4\right.\nonumber\\
&&\left.+\left(\frac{v_\Delta}{v_F}\right)^2\sin^2\theta\left[3(C_1-C_3)x^2+
(C_3-C_2)\cos^2\theta\right]\right\},\\
C_n&=&\frac{8v_\Delta}{N_f\pi^3v_F}\int^{\infty}_{-\infty}dx\int^{2\pi}_{0}
d\theta\frac{x^2\mathcal {G}(x,\theta)}{\left[x^2+\cos^2\theta+\left(\frac{v_\Delta}
{v_F}\right)^2\sin^2\theta\right]^3}\nonumber\\
&&\times\left[x^2-3\cos^2\theta-3\left(\frac{v_\Delta}{v_F}\right)^2\sin^2\theta\right],
\end{eqnarray}
with
\begin{eqnarray}
E_1&=&\frac{8\sqrt{2}v_\Delta}{\pi^{5/2}v_F^3}\int^{\infty}_{-\infty}dy \int^{2\pi}_{0}d\varphi
\frac{\mathcal {G}(y,\varphi)}{\left[y^2+\cos^2\varphi+\left(\frac{v_\Delta}{v_F}\right)^2
\sin^2\varphi\right]^4}\nonumber\\
&&\times\left\{\left[y^2+\cos^2\varphi-\left(\frac{v_\Delta}{v_F}\right)^2
\sin^2\varphi\right]\left[y^2+\left(\frac{v_\Delta}{v_F}\right)^2
\sin^2\varphi+\cos^2\varphi\right]\right.\nonumber\\
&&\left.-2y^2\left[y^2+\cos^2\varphi-3\left(\frac{v_\Delta}
{v_F}\right)^2\sin^2\varphi\right]\right\},\\
E_2&=&\frac{8\sqrt{2}v_\Delta}{\pi^{5/2}v_F^3}\int^{\infty}_{-\infty}dy \int^{2\pi}_{0}
d\varphi\frac{(\cos^2\varphi-y^2)\mathcal {G}(y,\varphi)}{\left[y^2+\cos^2\varphi
+\left(\frac{v_\Delta}{v_F}\right)^2\sin^2\varphi\right]^3},\\
E_3&=&\frac{8\sqrt{2}v_\Delta}{\pi^{5/2}v_F^3}\int^{\infty}_{-\infty}dy
\int^{2\pi}_{0}d\varphi\frac{\mathcal{G}(y,\varphi)}{\left[y^2+\cos^2\varphi
+\left(\frac{v_\Delta}{v_F}\right)^2\sin^2\varphi\right]^4}\nonumber\\
&&\times\left\{\left[\left(\frac{v_\Delta}{v_F}\right)^2\sin^2\varphi-y^2\right]
\left[\left(\frac{v_\Delta}{v_F}\right)^2\sin^2\varphi+\cos^2\varphi+y^2\right]\right.\nonumber\\
&&\left.+2\left(\frac{v_\Delta}{v_F}\right)^2\sin^2\varphi\left[3y^2-\cos^2\varphi -
\left(\frac{v_\Delta}{v_F}\right)^2\sin^2\varphi\right]\right\}.
\end{eqnarray}
\end{widetext}

\end{document}